\documentclass[12pt,preprint]{aastex}
\usepackage{longtable}

\makeatletter
\def\jnl@aj{AJ}
\ifx\revtex@jnl\jnl@aj\fi
\makeatother

\slugcomment{To be submitted to \aj}

\shorttitle{Variable stars in COROT LRa1 field}

\shortauthors{Kabath et al.}

\begin{document}

\title{\begin{center}Characterization of CoRoT target fields with BEST: Identification of periodic variable stars in the LRa1 field\end{center}}

\author{Kabath P., Eigm\"uller P., Erikson A., Hedelt P., Kirste S., von Paris P.,Rauer H.\altaffilmark{1}, Renner S., Titz R., Wiese T.}

\affil{Institut f\"ur Planetenforschung, Deutsches Zentrum f\"ur Luft- und Raumfahrt, 12489 Berlin, Germany}

\and

\author{Karoff C.}

\affil{Department of Physics and Astronomy, University of Aarhus, Ny
Munkegade, Building 1520, DK-8000 Aarhus C, Denmark}

\email{petr.kabath@dlr.de}

\altaffiltext{1}{Also: Zentrum
f\"ur Astronomie und Astrophysik, Technische Universit\"at Berlin,
Germany}
\begin{abstract}

In this paper we report on observations of the CoRoT LRa1 field with the Berlin
Exoplanet Search Telescope (BEST). The current paper is part of the series of papers describing the results of our stellar variability survey. BEST is a small aperture telescope with a
wide field-of-view (FOV). It is dedicated to search for stellar variability within the target fields of the CoRoT space mission to aid in minimizing false-alarm rates and identify potential targets for
additional science. The LRa1 field is CoRoT's third observed field and the second long run field located in the galactic anticenter direction. We observed the LRa1 stellar field on $23$ nights between November and March $2005/2006$. From $6099$ stars marked as variable, $39$ were classified as periodic variable stars and $27$ of them are within the CoRoT FOV. We also confirmed the variability for $4$ stars listed in GCVS catalog.  

\end{abstract}


\section{Introduction}

In previous articles belonging to a series dedicated to the variability survey in the CoRoT observational fields with BEST we presented the results of our survey on variable stars in the CoRoT IRa1~\cite{Kabath07} and LRc1~\cite{Karoff07} observational fields. The fields were observed with BEST~\cite{Rauer04}, a small aperture telescope with a wide field-of-view (FOV) developed and operated by the Institut f\"{u}r Planetenforschung of Deutsches Zentrum f\"ur Luft- und Raumfahrt (DLR). As the BEST magnitude range covers approximately that of the CoRoT it is thus well suited for ground based support of the CoRoT space mission~\cite{Baglin98}. 

CoRoT was launched in December $2006$. The scientific goals of the mission are observations of selected stellar fields in order to find transiting extrasolar planets and to perform astroseismology of selected stars. The duration of the mission is planned to be $2.5$ years with $4$ long ($150$ days), $4$ short (up to $30$ days) and one initial ($30$ days) run. The CCD camera of CoRoT is divided into four segments from which two are dedicated to a transiting extrasolar planets survey and the other two to the asteroseismic survey. The total FOV covers $2.7^\circ \times 3.05^\circ$. The point spread function (PSF) of CoRoT's extrasolar planet survey is about $80$ pixels for a $13$ mag K2 star and the PSF of the astroseismology field is about $410$ pixels for a solar type star at magnitude $5.7$ see~\cite{Boisnard2006}. The prisms mounted in front of CoRoT's CCD provide colour information on observed objects which can help to determine the type of variability of the central star. 

The BEST survey telescope system is used to discover and characterize variable stars in the CoRoT fields prior to CoRoT observations. The observational data can also be used as a complimentary information to the incoming CoRoT data. In addition BEST observations should also point out potentially interesting objects for the CoRoT additional science programs, such as binaries, $\delta$ Scuti stars etc.. The advantage of our observations is the long time line which can provide extended lightcurves for several thousands of stars in the comparable magnitude range as CoRoT data and thus better understanding of e.g. potential variation of the lightcurve. 

The resulting information about the variability in the stellar fields observed by BEST will be provided to the scientific community via the BEST archive linked to CoRoT's EXODAT~\cite{Deleuil2006} database. 

\section{BEST setup and observations}

The BEST telescope was described in detail in earlier papers~\cite{Rauer04}. Thus, we just briefly summarize the main parameters of the system here. Since the winter season $2005$ BEST operates at the Observatoire de Haute-Provence (OHP), France. A remote control operational mode from Berlin was implemented during the summer season $2006$. BEST is a Schmidt-Cassegrain telescope system with an aperture of $19.5$ cm. The $3.1^\circ \times 3.1^\circ$ FOV is monitored with a peltier cooled $2048\times 2048$ Apogee AP-10 CCD. The pixel scale is $5.5''$ per pixel. The digital resolution of the CCD is $14$ bit with a saturation level of $16384$ ADU and a short readout time of $9$ sec.   

The observational run consists of a sequence of $40$ sec and $240$ sec images followed by a set of appropriate dark frames and bias frames approximately every $20$ minutes. The whole sequence takes about $40$ minutes and is repeated as long as the target field is visible. After each sequence, a set of bias frames is taken. The telescope is not equipped with any filter but the bandwidth corresponds approximately to a wide $R$-bandpass filter. 

In this paper we present the results obtained from the reduction of $434$ images of $240$ sec exposures times each because the resulting magnitude range overlaps with the CoRoT magnitude range.    

The CoRoT LRa1 field is located at $\alpha=06^h 46^{m} 53^{s}$ and $\delta=-00^\circ 12' 00''$. The approximate coordinates of the centre of the BEST FOV are $\alpha=06^h 46^{m} 24^{s}$ and $\delta=-01^\circ 54' 00''$. The LRa1 was observed between $21^{st}$ November $2005$ and $6^{th}$ March $2006$. The total time baseline of our observations spreads over $104.937$ days. In general no data were obtained at nights with bad weather conditions and three days before and three days after the full Moon.

\section{Data processing}

The data reduction and analysis was described in \cite{Karoff07} and \cite{Kabath07} in detail. The standard photometric calibration on raw images was performed. Then the image subtraction routine ISIS~\cite{Alard98} was applied on the calibrated data set in the following steps. A reference frame was chosen to find a reference template for the coordinate system for all images. Five images with the best seeing were then chosen from the data set and were stacked together. Kernels were found on the subsectors of the stacked reference frame and they were subtracted from the point spread functions (PSF) of the stars in those sectors. We did not make use of an ISIS photometry routine for the further reduction. Our photometry routine using a unit-weighted aperture photometry was used on the subtracted images and on the reference image with an aperture radius of $7$ pixels. The differential magnitudes were estimated in respect to the reference frame. In the next step a cubic spline function within a MATCH routine~\cite{Valdes} was used to find a transformation between internal frame coordinates and right ascension and declination. For this purpose the $1500$ brightest stars from the BEST data set were compared with the USNO-A2.0 catalogue~\footnote{Monet}. In the last step, the zero offset variations between nights were corrected for all frames.  

\begin{figure}
\begin{center}
\includegraphics[angle=90, scale=.30, width=14cm]{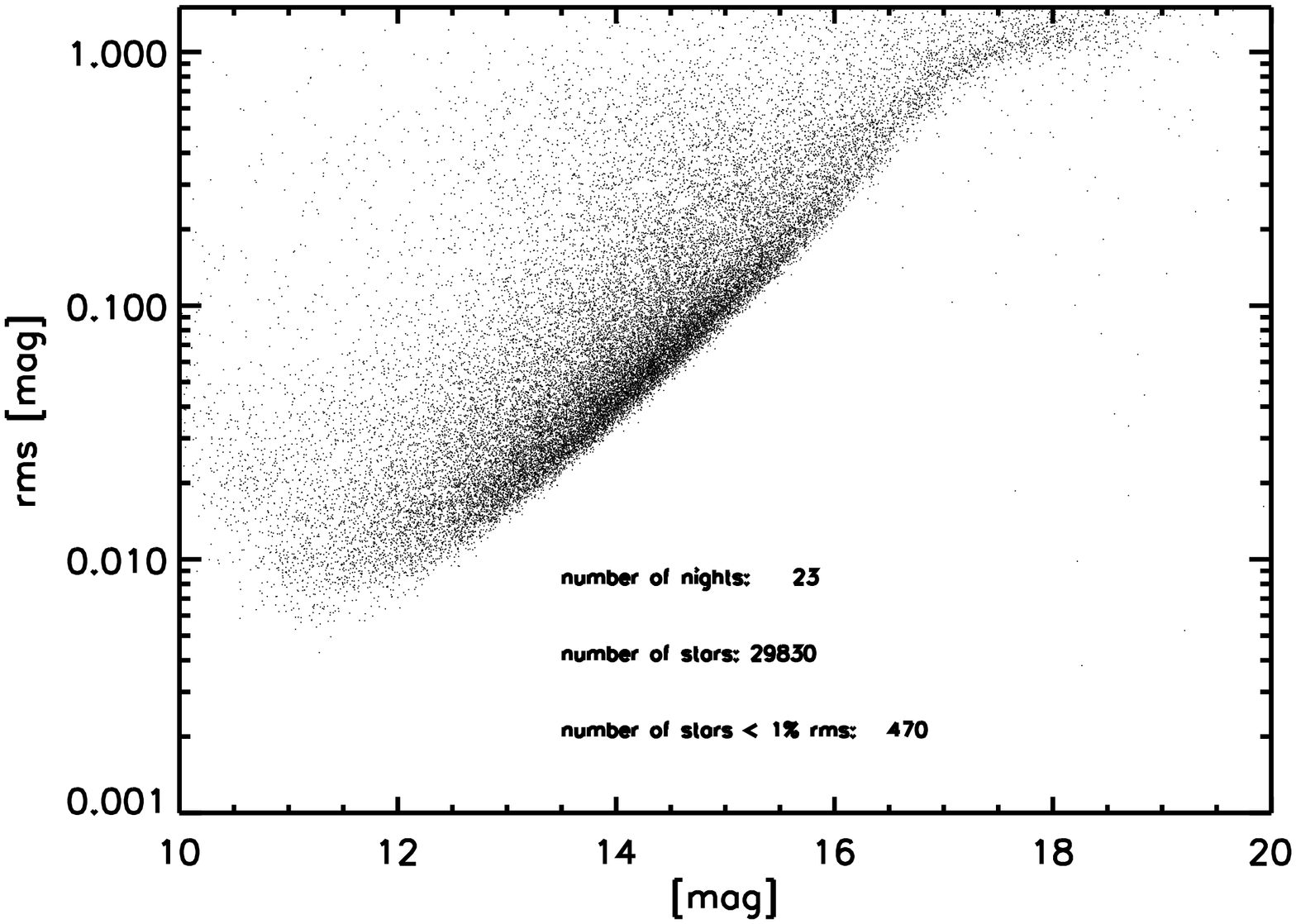}
\caption{The rms noise level of stellar lightcurves for the whole data set of $23$ nights over magnitude for the LRa01 observing campaign.\label{rms}}
\end{center}
\end{figure}

 An additional correction of photometry based on the algorithm by Tamuz~\cite{Tamuz05} was also performed on the reduced data. The photometry correction algorithm should remove the remaining systematic errors e.g. due to atmospheric extinction. The RMS plot for the whole campaign is shown in Figure~\ref{rms}. We propose that we do not aim for the milimagnitude precision absolute photometry.   

\section{Variable stars}

\subsection{Criterion for variability}

To be able to select variable stars from the huge data set we used a modification of the Stetson's $j$-variability index~\cite{Stetson1996} according to Zhang~\cite{Zhang2003}. Stars which were suspected from variability were chosen among the stars having $j > 2$. In result $6099$ stars from our data set of $29830$ detected stars sattisfied our empirical variability criterion. The distribution of the variable stars marked with an index $j$ over magnitude range is shown in Fig. 3 for all observed stars. However the results presented in more detail in the following section refer only to the periodic variable stars.

\subsection{Detected periodic variable stars}

All lightcurves of suspected variable stars in the LRa1 field were searched for periodicity with the method introduced by Schwarzenberg-Czerny (1996), which fits a set of periodic orthogonal polynomials to the observational
data sets. As a criterion for the quality of the fit a variance statistics is used. We examined visually all possible variables with reliability of the fit higher than $0.9$. 

Reasons for rejection were a period very near to one day, or to multiples of one day and when the lightcurve did not show clear variability upon the visual inspection. 

We identified in total $43$ periodic variable stars and one longperiodic star was identified after comparison of the BEST field with the SIMBAD catalog. The detailed characteristics of the identified stars are listed in Table 1. The stars which are located in CoRoT fields are marked with an asterisk. In Fig. 2 the positions of the the identified stars in the BEST field is shown.

The classification is as in previous papers based on the GCVS catalog (see Sterken $\&$ Jaschek, 1996). Our data set is limited and no color neither any spectral information is available. The groups used for the classification are: CEP, DSCT, ELL, EA, EB and EW and they will be described later on in the following paragraphs.

Clear separation in the classification is between pulsating, eclipsing and rotating variable stars. The pulsating stars are divided into two subgroups DSCT and CEP based on the period. Stars with a period below $0.3$ days are classified as DSCT ($\delta$ Scuti type) and the stars with longer periods are classified as CEP (Cepheid star type). The criterion for the rotating stars is the unequality of the maxima/minima of the lightcurves. We identified newly $21$ stars belonging to the CEP group, $3$ stars belonging to the DSCT group and $4$ ELL stars. The lightcurve of the star $211$ is probably affected with a flux from the nearby star. 

The eclipsing stars were divided into three groups EA (Algol type eclipsing stars), EB ($\beta$ Lyrae type eclipsing stars), EW (W Uma type stars) depending on the type of the lightcurve. EA class shows constant lightcurves between eclipses, EB class is continuously varying between eclipses and the stars from the EW class show an equal depth and the periods shorter than $1$ day. In our data sets we identified newly $11$ eclipsing stars. Folded lightcurves for all types of newly detected stars are shown in Fig. 4.

The detected periodic variable stars in the observational field were compared also with the GCVS catalogue via SIMBAD. The BEST data set contains the lightcurves of five already known variable stars. Two of them, GU Mon (EW W Uma type) and DD Mon (EB $\beta$ Lyr type), correspond to BEST stars nr. $2361$ and nr. $3550$ respectively. Our periods of $0.8960$ days (GU Mon) and $0.5684$ days (DD Mon) correspond well with the periods listed in the GCVS ($0.896681490$ and $0.5680119300$ days respectively)~\footnote{VIZIER}. The star V 404 Mon (EA/DCEP type), in our data set star nr. $4559$ shows a period of $2.4452$ days which also corresponds as well with the period listed in GCVS ($2.445205300$ days). We were not able to identify the periodicity for the V 501 Mon (EA/DCEP type), in our data set star nr. $5239$ because we did not observe any extrema of the lightcurve due to the relatively long period of $7.0211710000$ days~\footnote{VIZIER} with respect to our duty cycle coverage. In addition a lightcurve for star CD Mon, in BEST data set nr. $25207$ is also present in our archive. CD Mon is a Mira type star with a period of $268.1$ days. Thus we are not able to confirm the period because of our duty cycle, but the partial lightcurve is available in our archive. The folded lightcurves for the stars GU Mon, DD Mon, V501 and V 404 Mon and a lightcurve of the CD Mon are shown in Fig. 5.

\section{Summary}

We performed photometry on CoRoT LRa1 field with BEST telescope to detect stellar variability. We identified $6099$ stars which were marked as variable and $44$ of them are showing a regular period. In total we detected $39$ new variable stars. The period ranges are usually between $0.1$ $<$ P $<$ $3$ days, however two new longperiodic stars with periods about $7$ and $21$ days were detected. The relatively small period range is given due to limited data set and due to duty cylce coverage which was disturbed by the bad weather period in December. We also confirm the period for the stars GU Mon, DD Mon and V 404 Mon which are already listed in the GCVS catalogue. The period for the GCVS star V 501 Mon could not be confirmed because no maxima/minima are present in our data set for that star. The star CD Mon is a longperiodic Mira type star and is also present in our data set. The newly found variable stars are currently observed within CoRoT's additional science programs. The information provided with our survey brings an additional information when analysing the CoRoT data and may avoid a false alarm for CoRoT transiting planet candidates. We gladly provide additional information about our data archive upon a request.

\acknowledgments

The authors gratefully acknowledge the support and assistance of the staff at Observatoire de Haute-Provence.

C. K. acknowledges
support from Danish AsteroSeismology Centre and Instrument Center
for Danish Astrophysics. The support to DLR staff in observational
tasks given by Berlin local amateur astronomers Susanne Hoffman,
Irene Berndt, Martin Dentel and Karsten Markus is also greatly
acknowledged. We made use of the SIMBAD and GCVS catalogs.

\clearpage

\begin{figure}
\begin{center}
\includegraphics[angle=90, scale=.45, width=12cm]{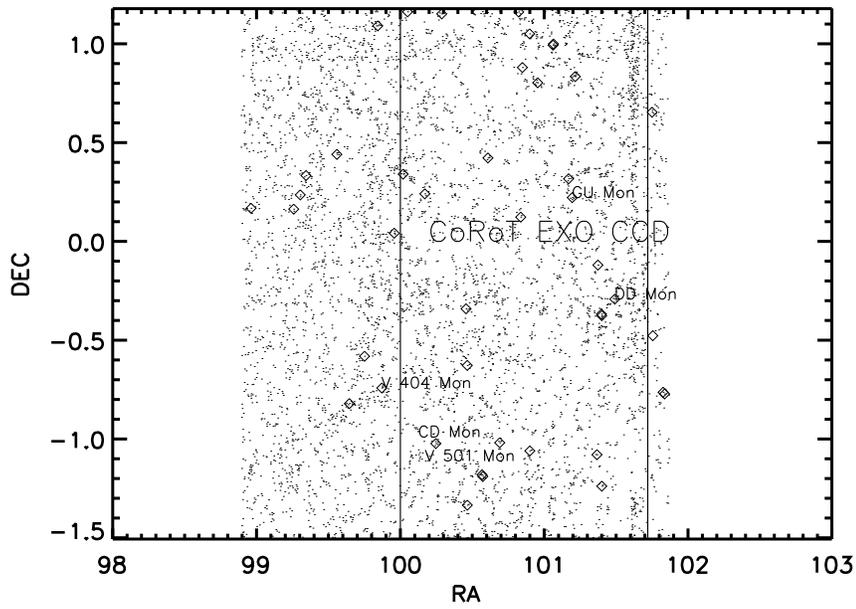}
\caption{The EXO-part of the CoRoT LRa1 field is indicated in the BEST field. The distribution of the variable stars found in the BEST FOV is shown. Dots show stars with $j>2$ and diamonds are periodic variable stars selected visually from the stars having $j>2$. \label{cam}}
\end{center}
\end{figure}

\clearpage

\begin{figure}
\begin{center}
\includegraphics[angle=90, scale=.45, width=14cm]{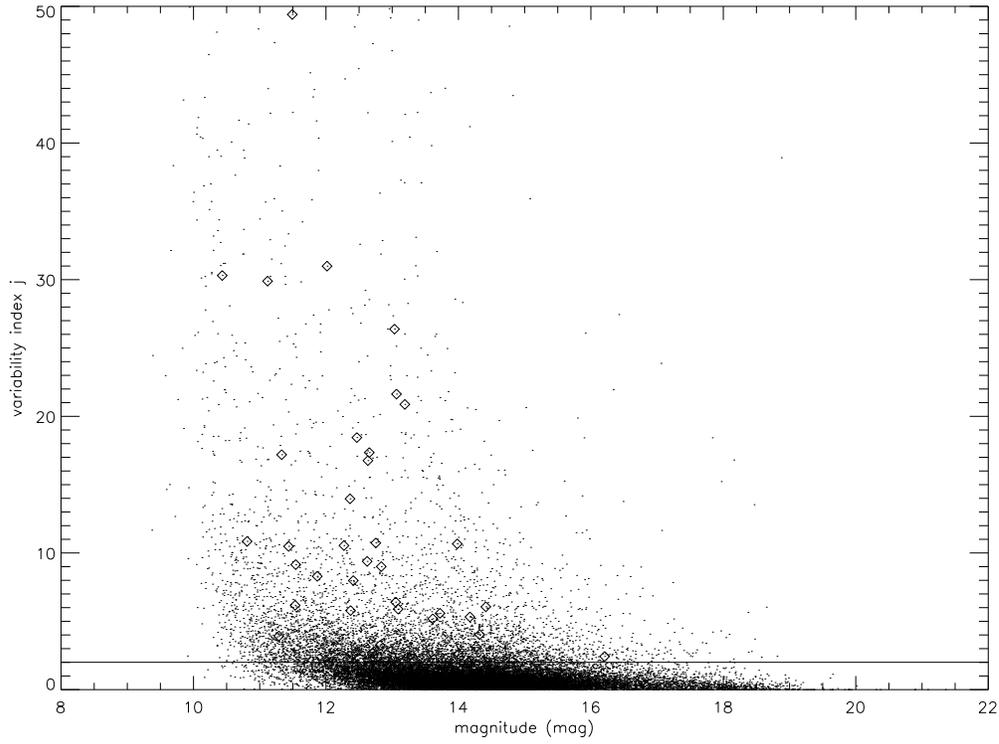}
\caption{Variability index $j$ of the sample plotted over magnitudes
of the stars. The line marks the limit of $j=2$ used to identify
suspected variable stars. The identified variable stars are represented with diamonds.\label{jindex}}
\end{center}
\end{figure}

\clearpage

\begin{figure}[ht]
\includegraphics[scale=.45]{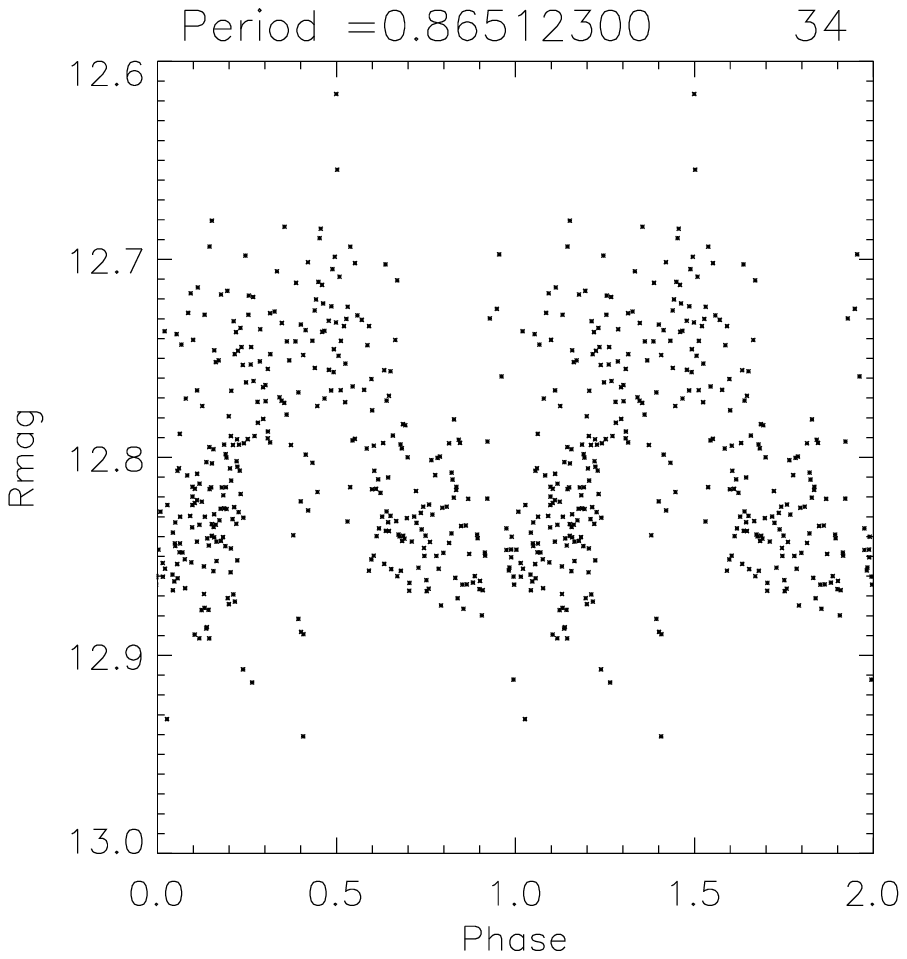}
\includegraphics[scale=.45]{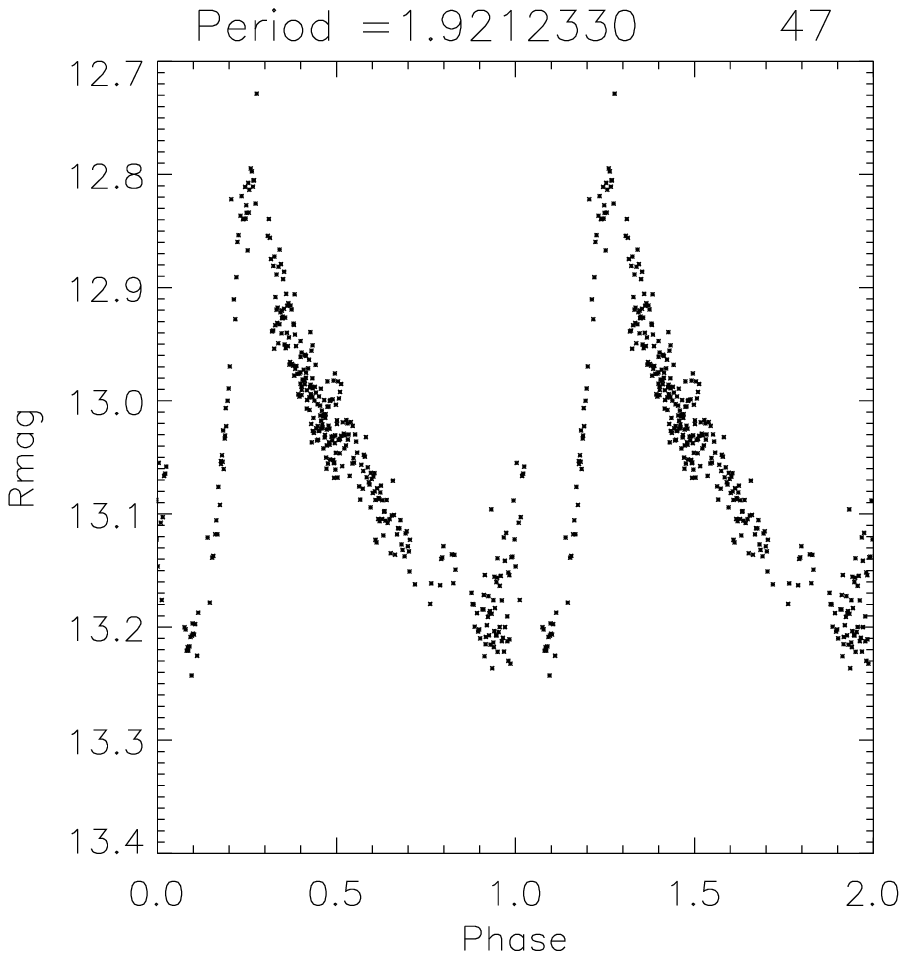}
\includegraphics[scale=.45]{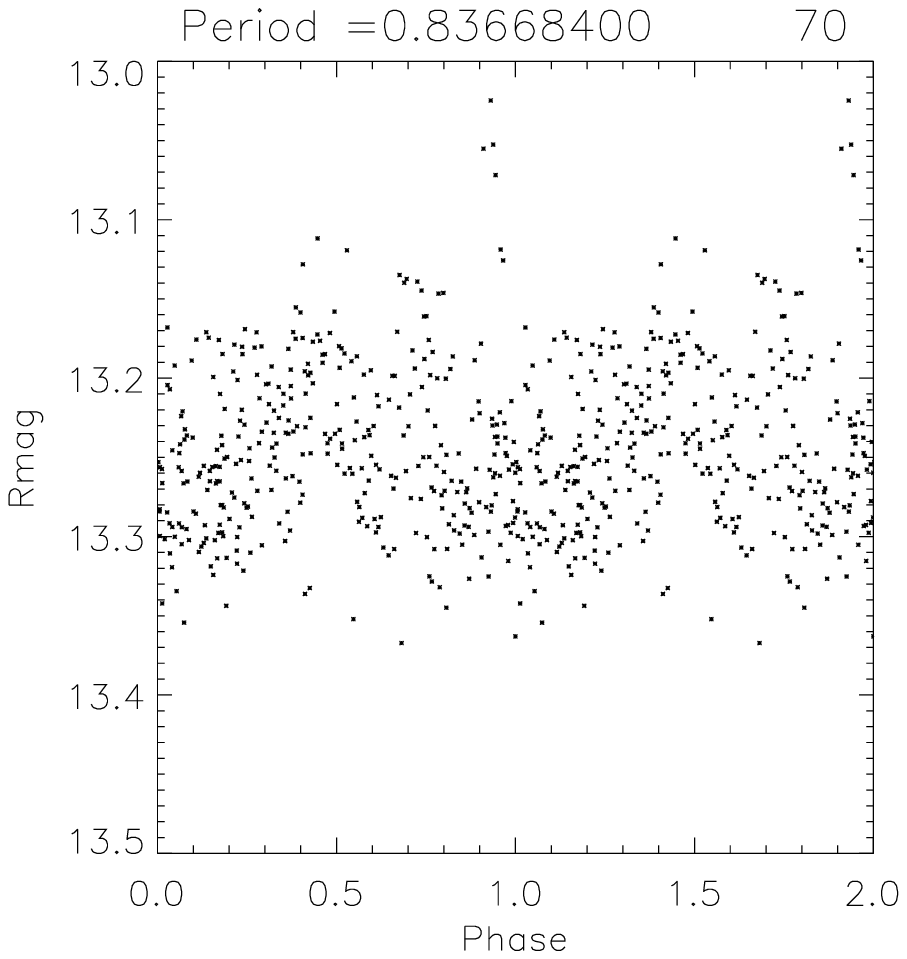}
\includegraphics[scale=.45]{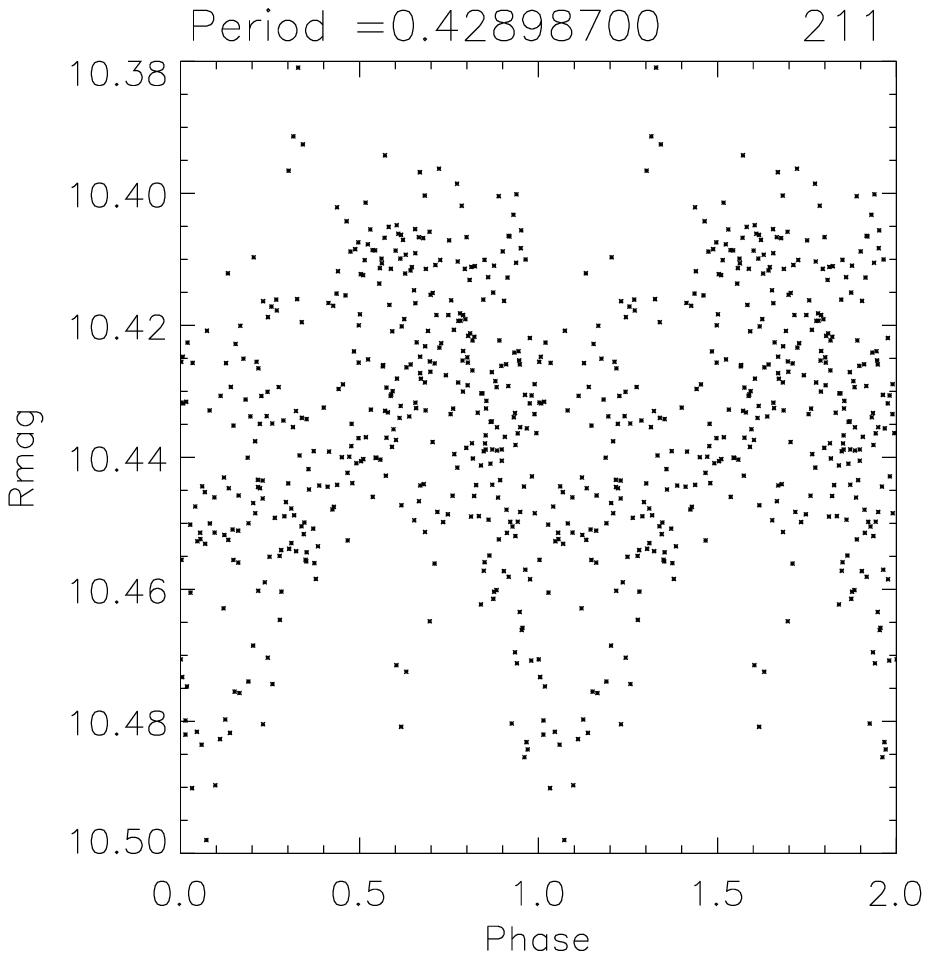}
\includegraphics[scale=.45]{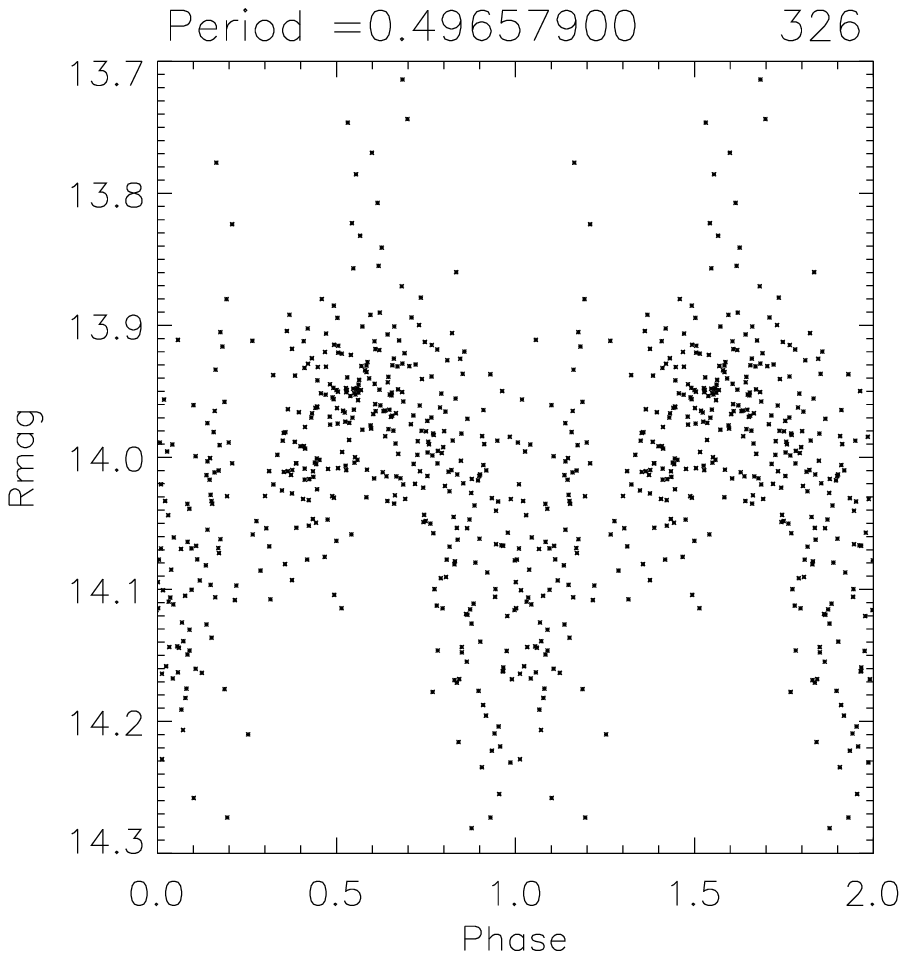}
\includegraphics[scale=.45]{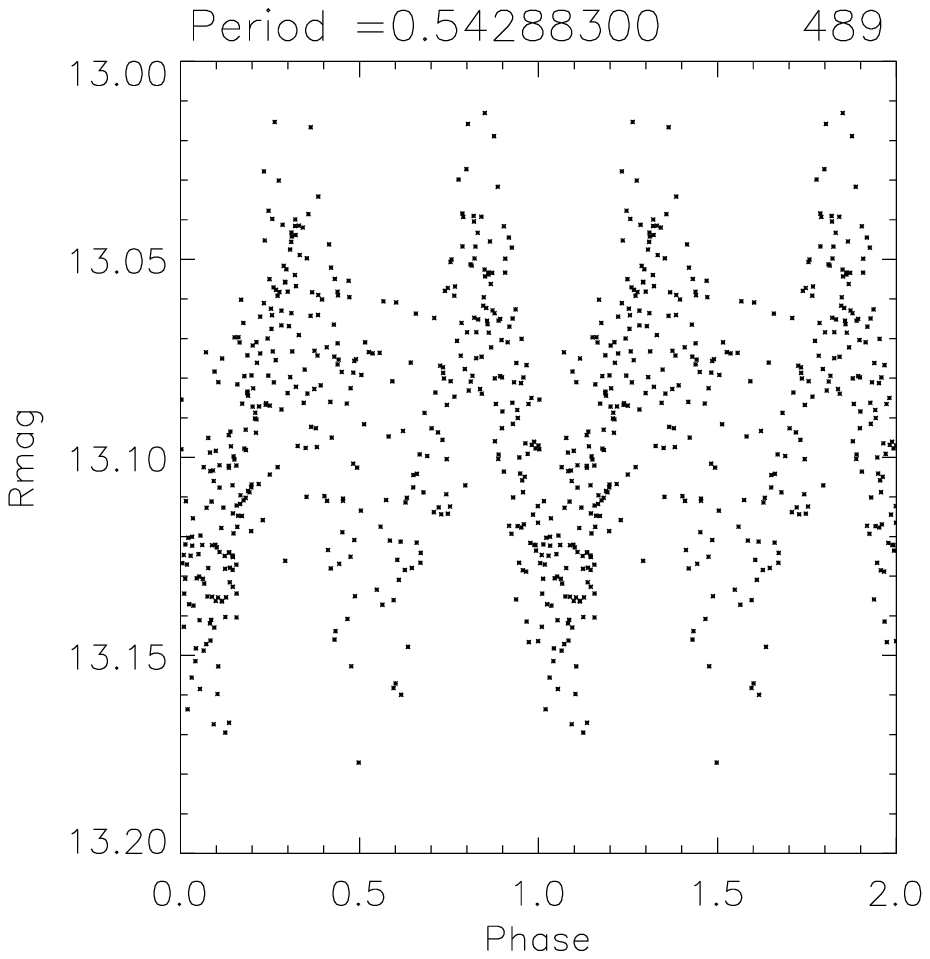}
\includegraphics[scale=.45]{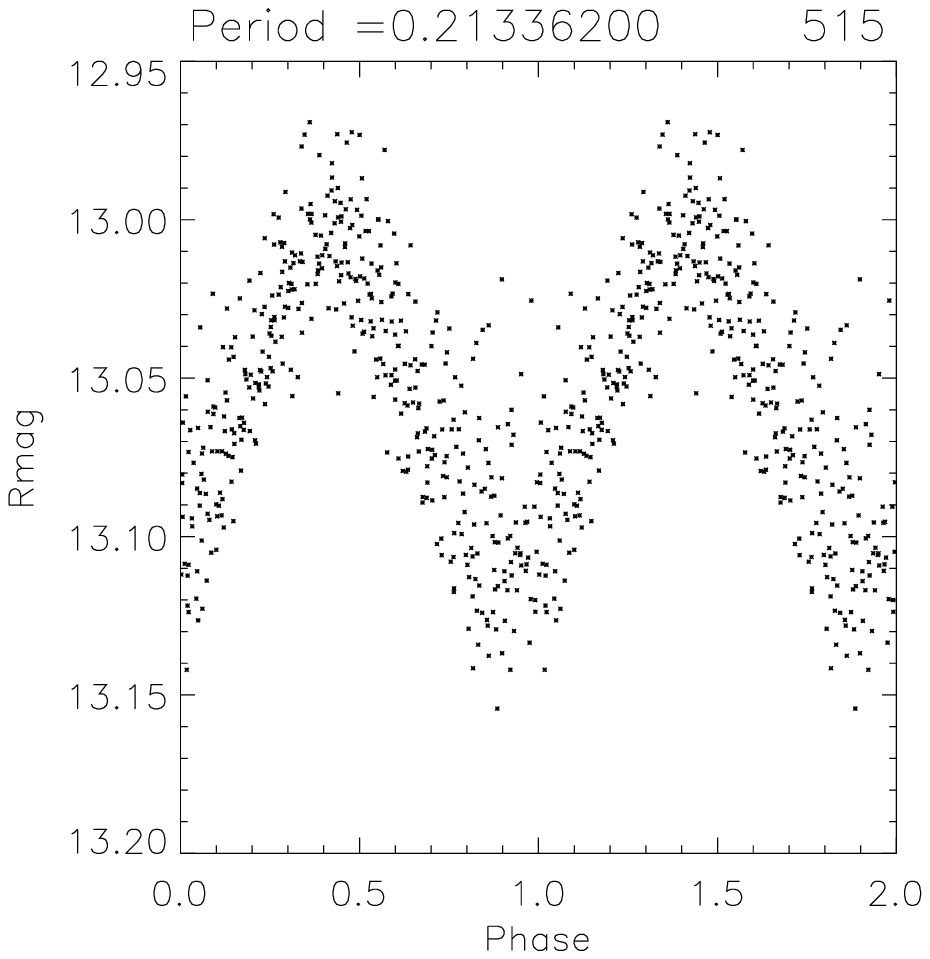}
\includegraphics[scale=.45]{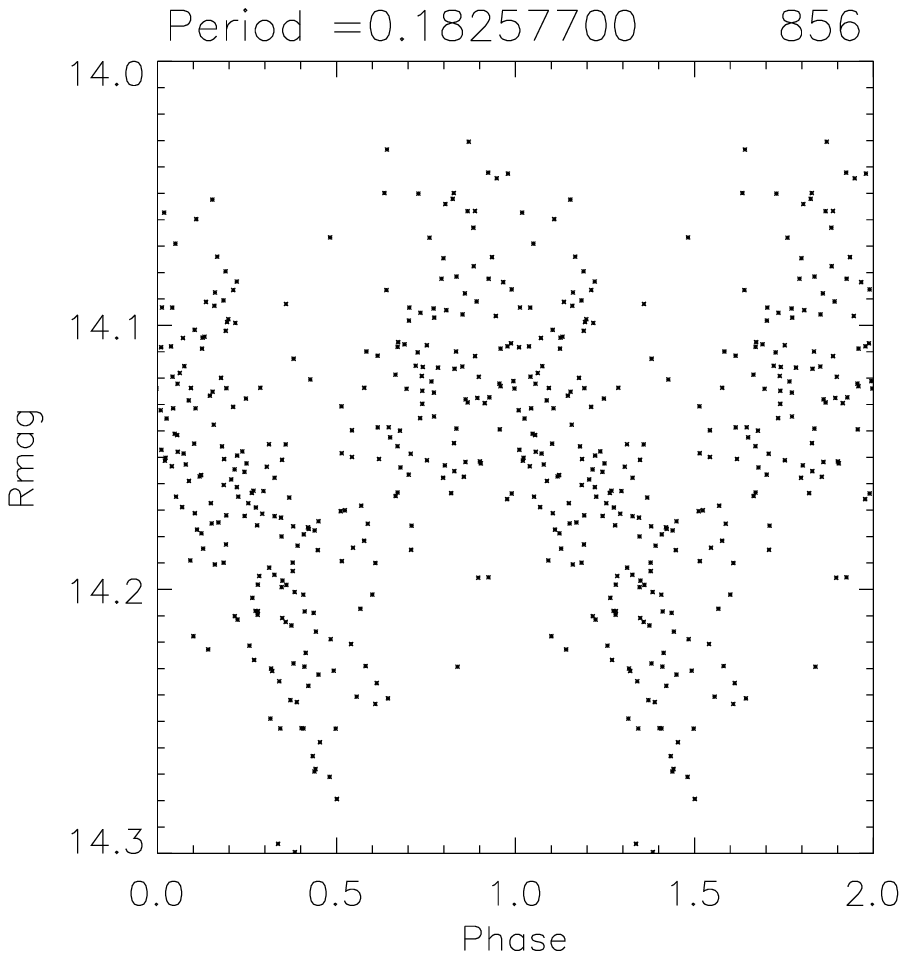}
\includegraphics[scale=.45]{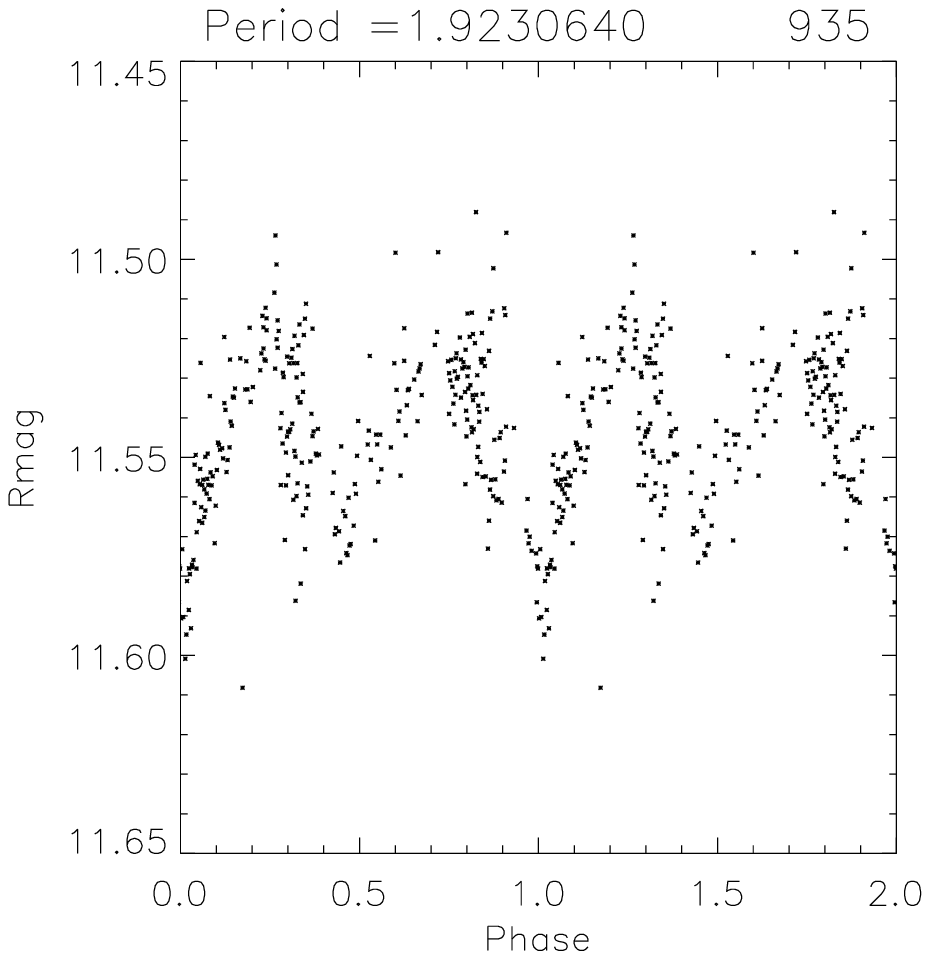}
\includegraphics[scale=.45]{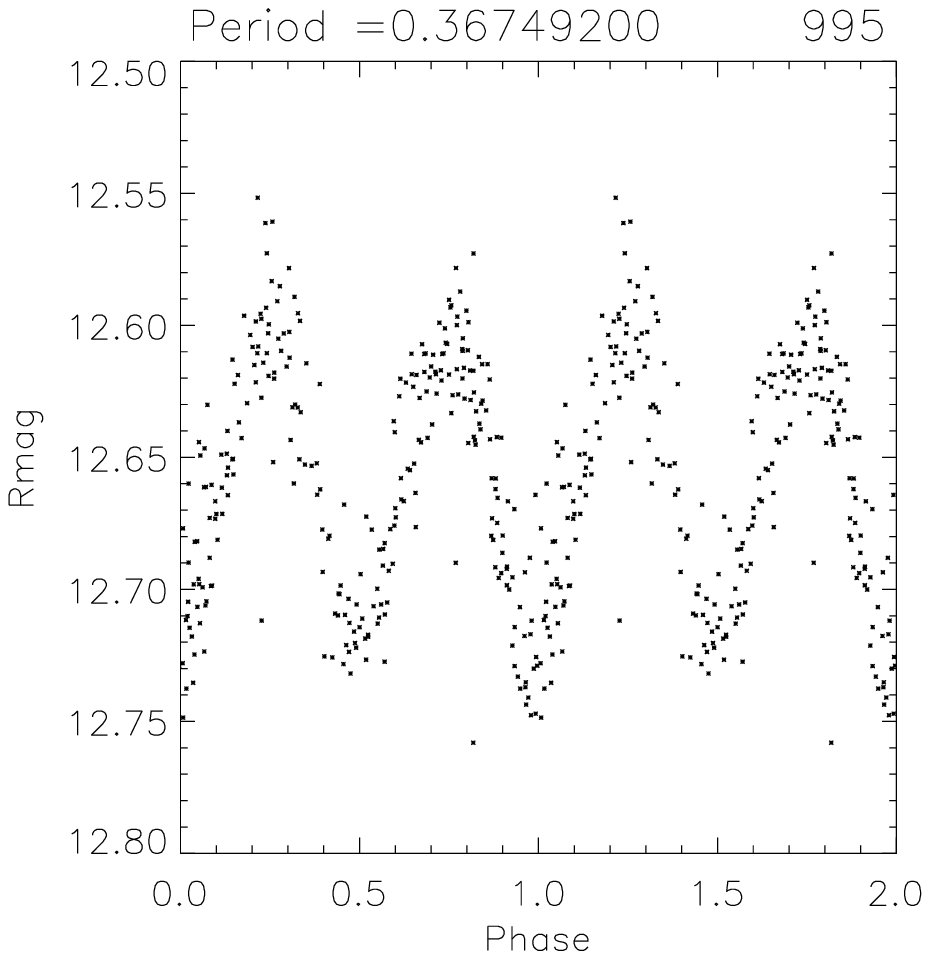}
\includegraphics[scale=.45]{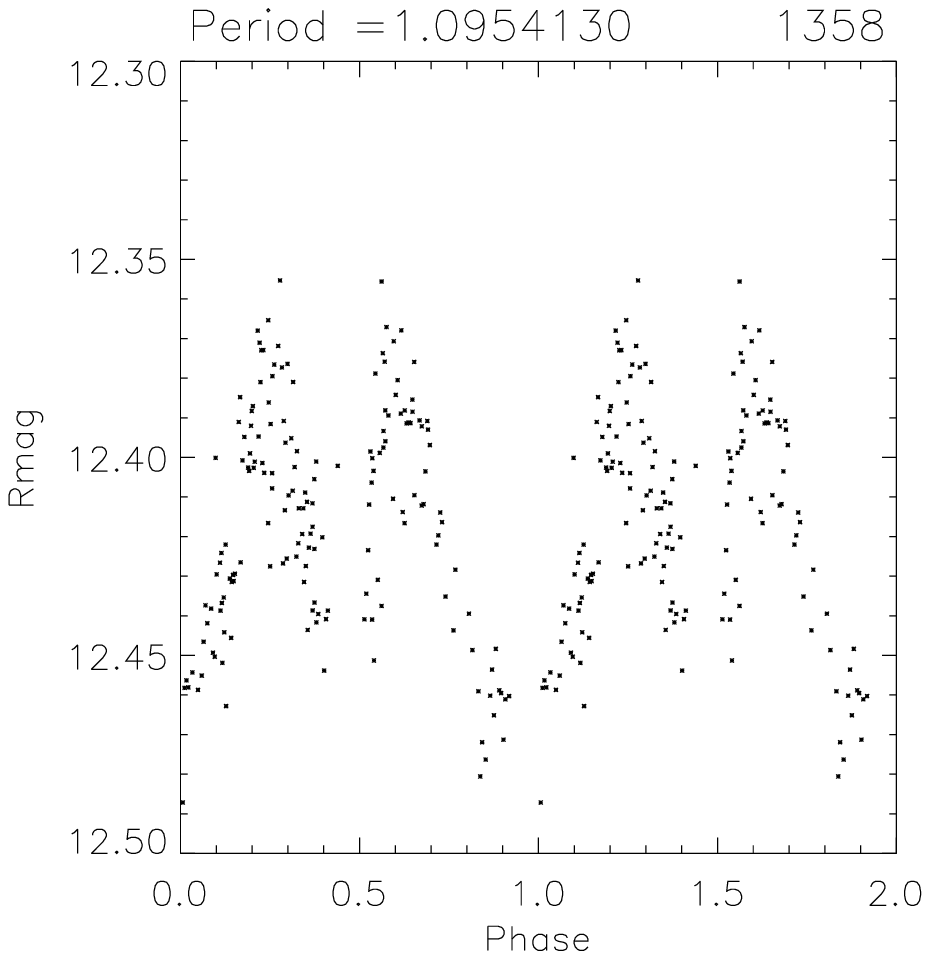}
\includegraphics[scale=.45]{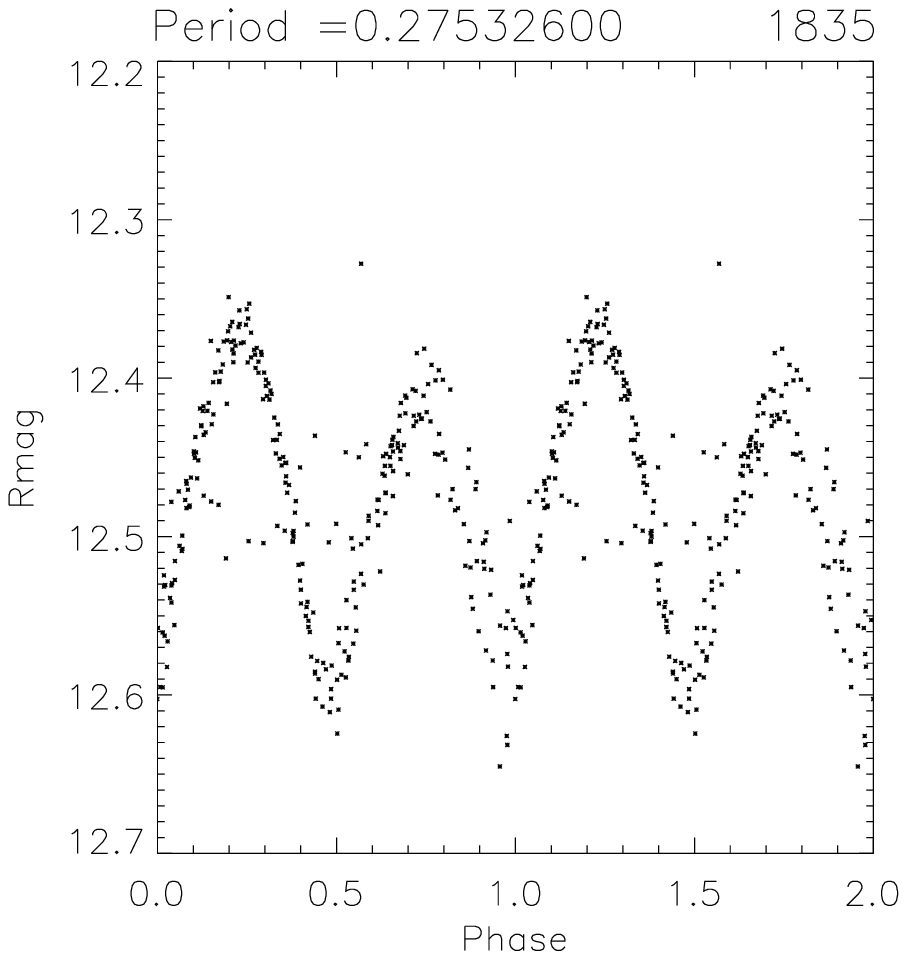}
\end{figure}
\clearpage
\begin{figure}[ht]
\includegraphics[scale=.45]{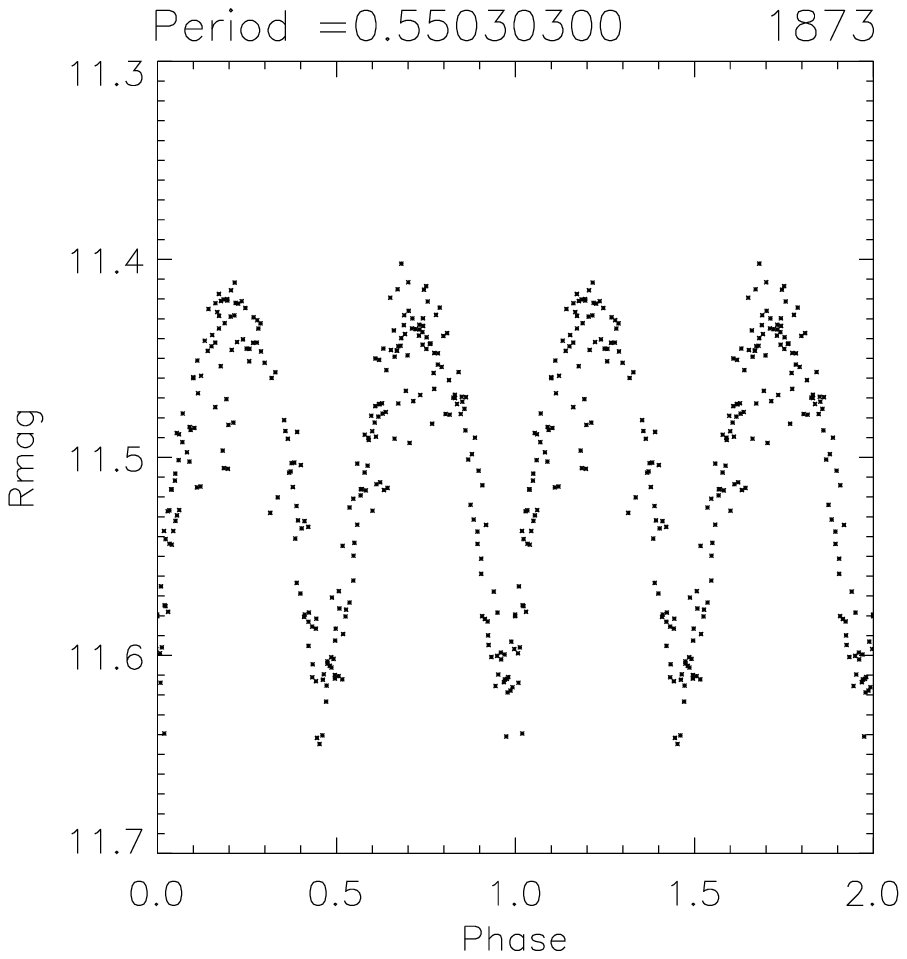}
\includegraphics[scale=.45]{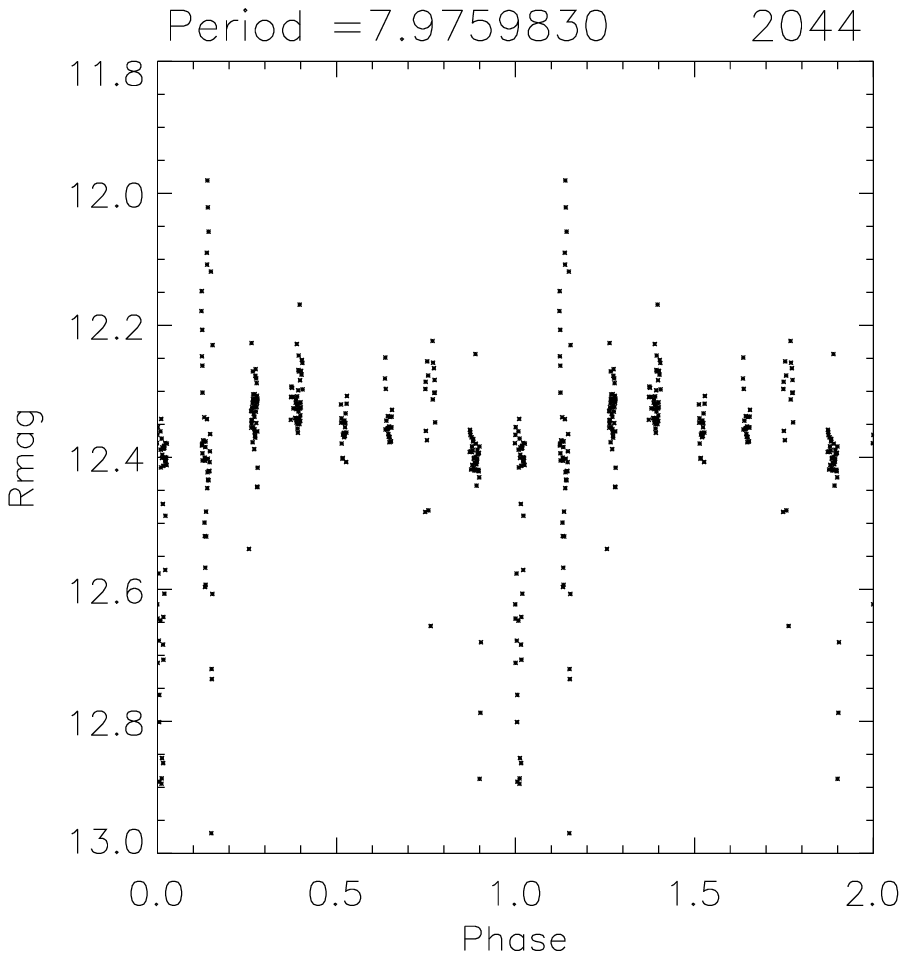}
\includegraphics[scale=.45]{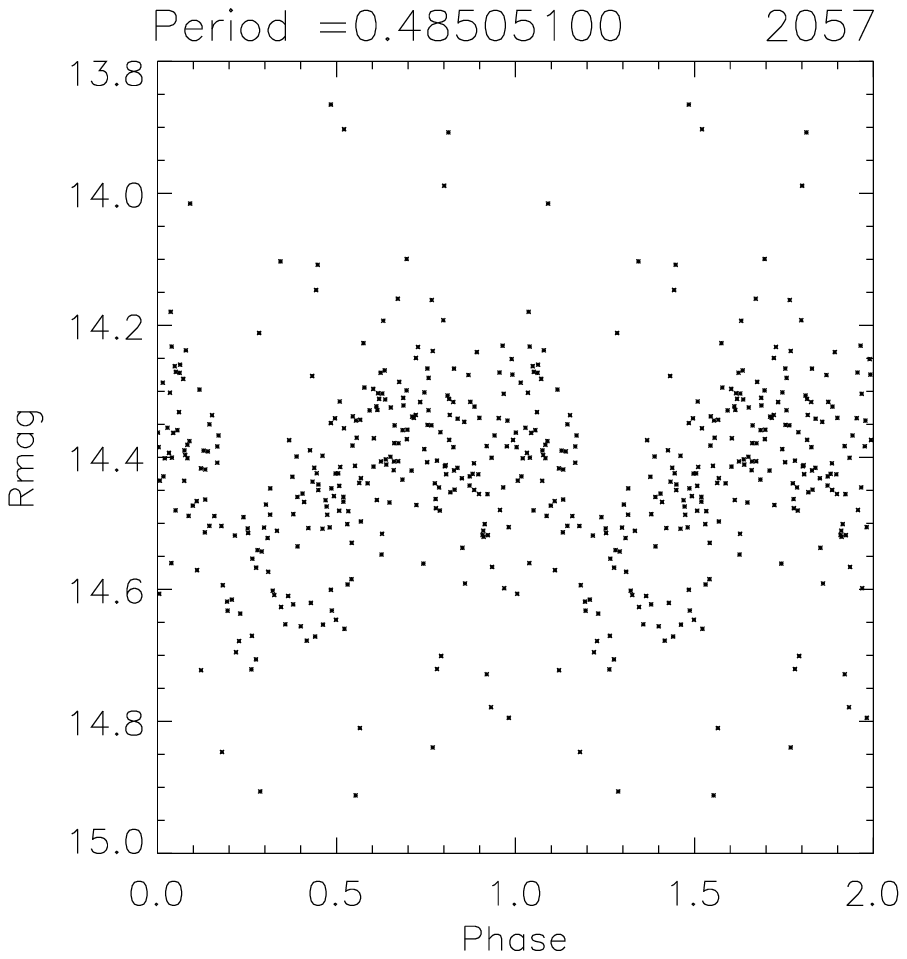}
\includegraphics[scale=.45]{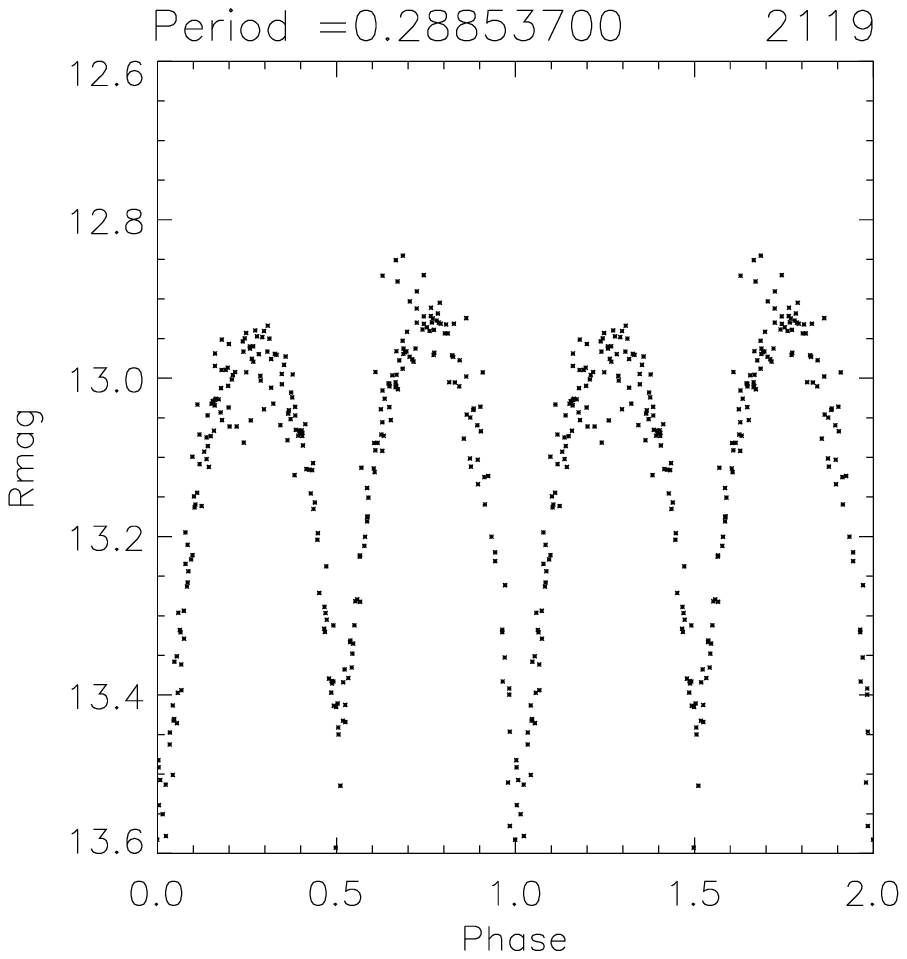}
\includegraphics[scale=.45]{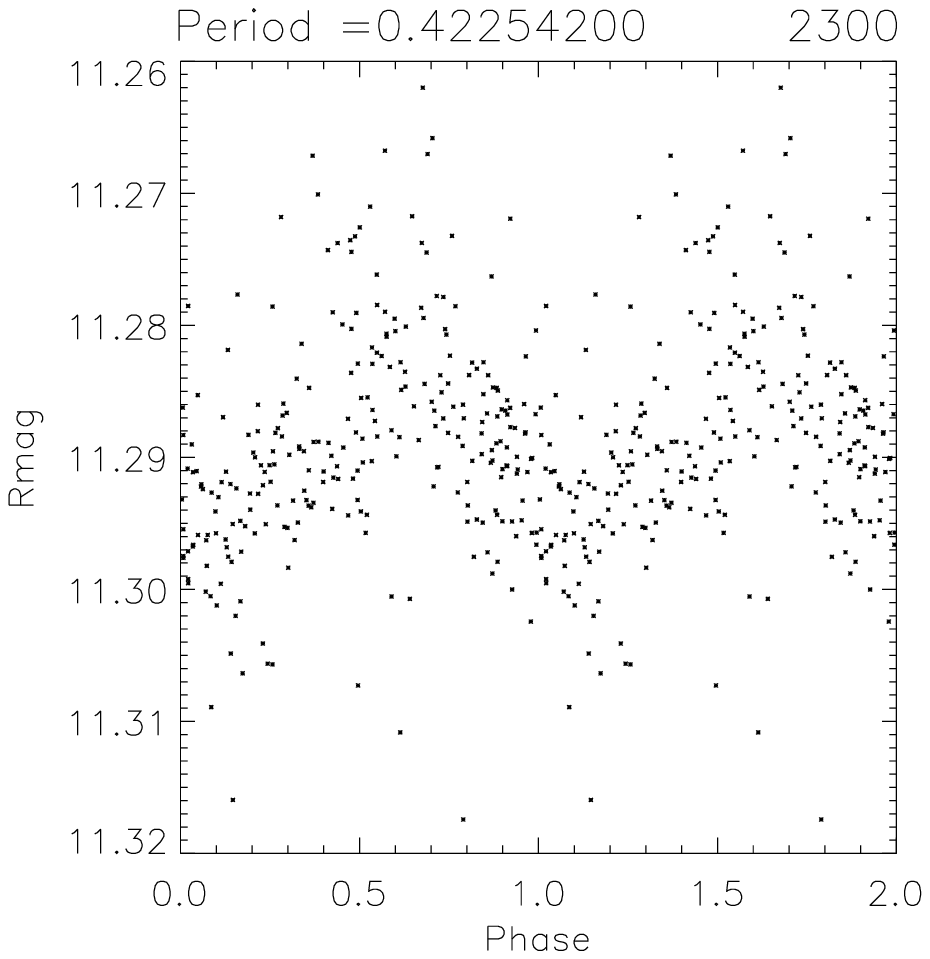}
\includegraphics[scale=.45]{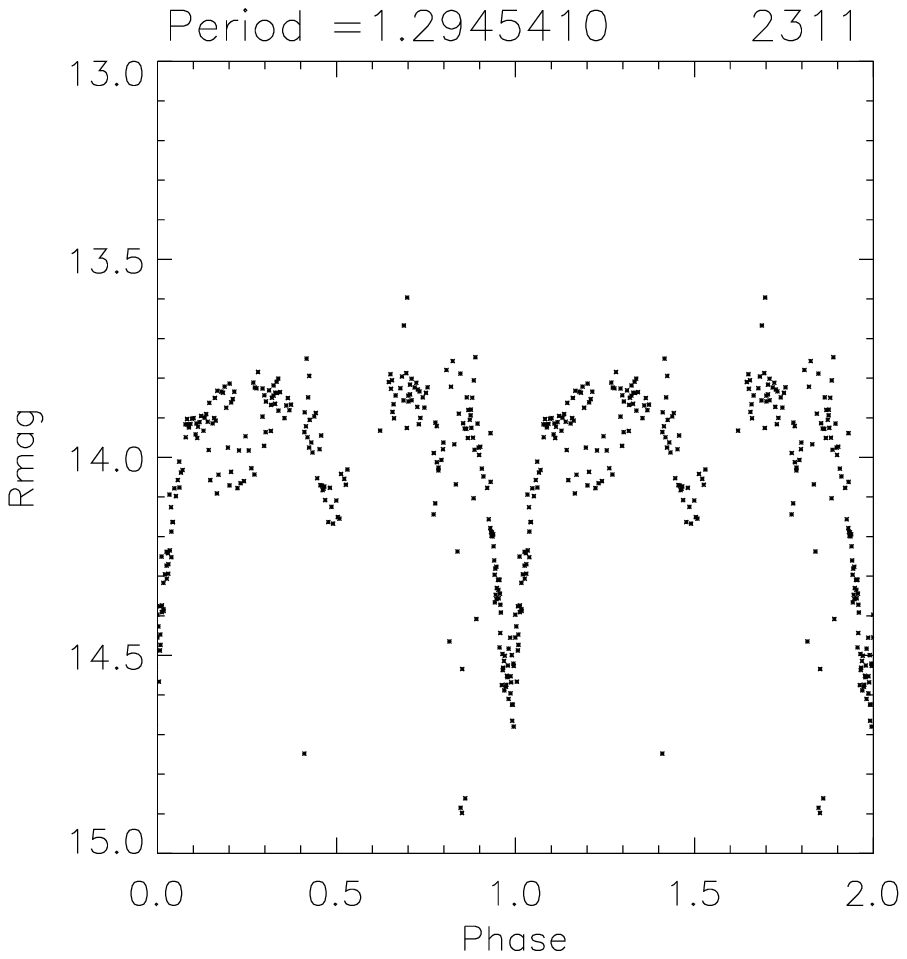}
\includegraphics[scale=.45]{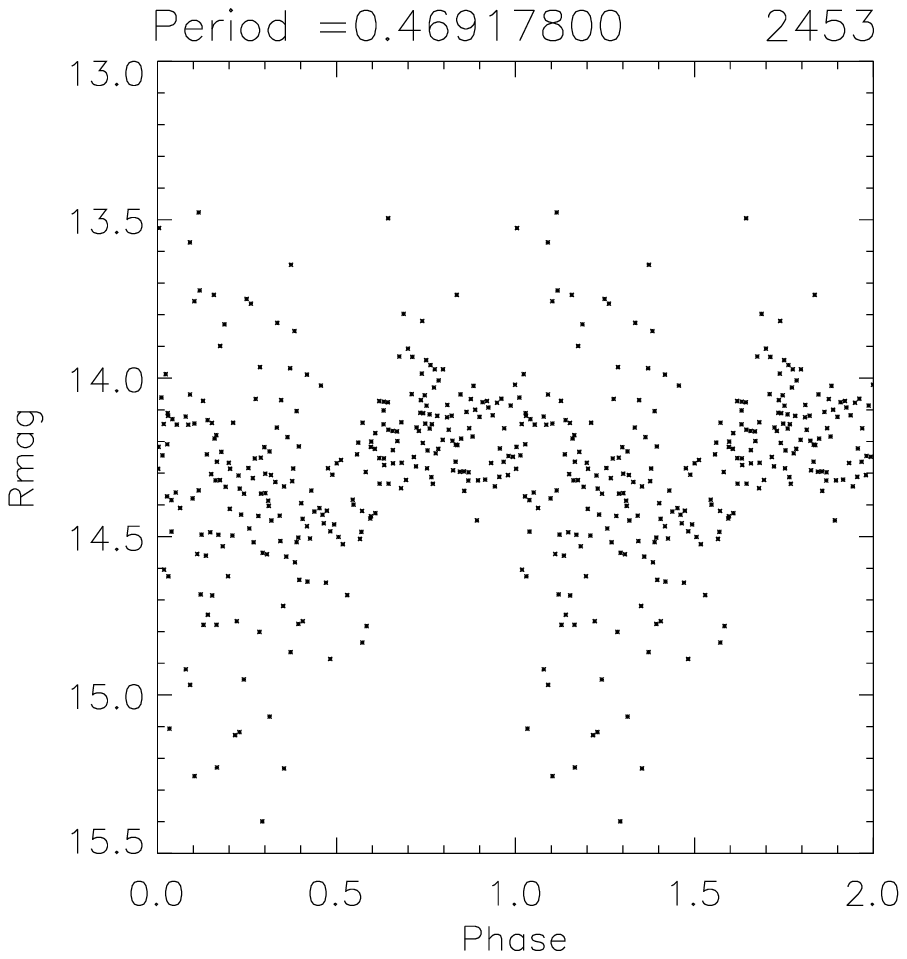}
\includegraphics[scale=.45]{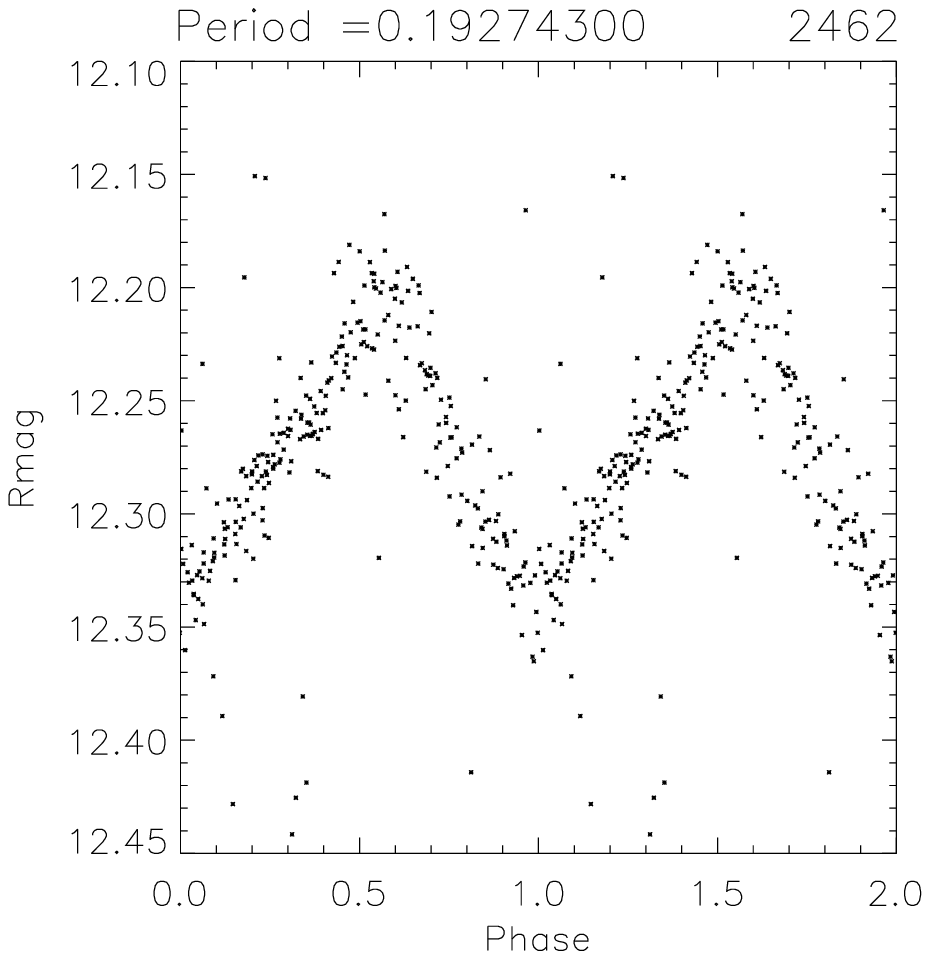}
\includegraphics[scale=.45]{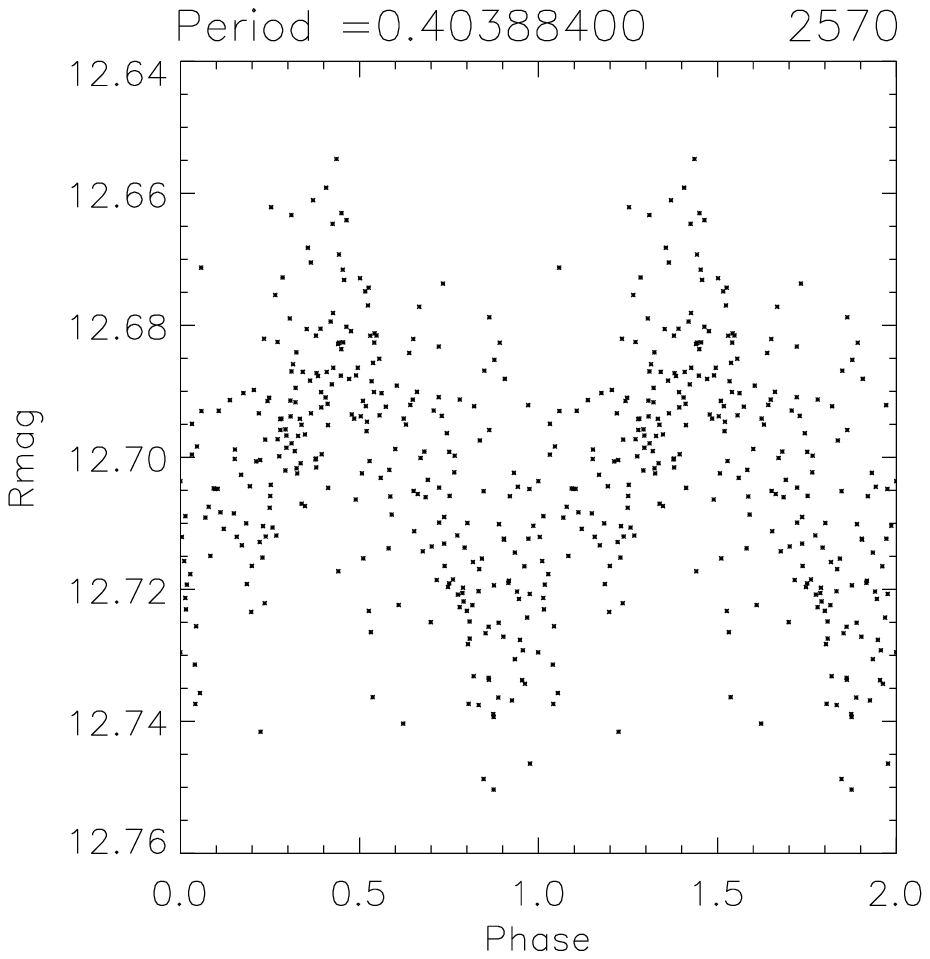}
\includegraphics[scale=.45]{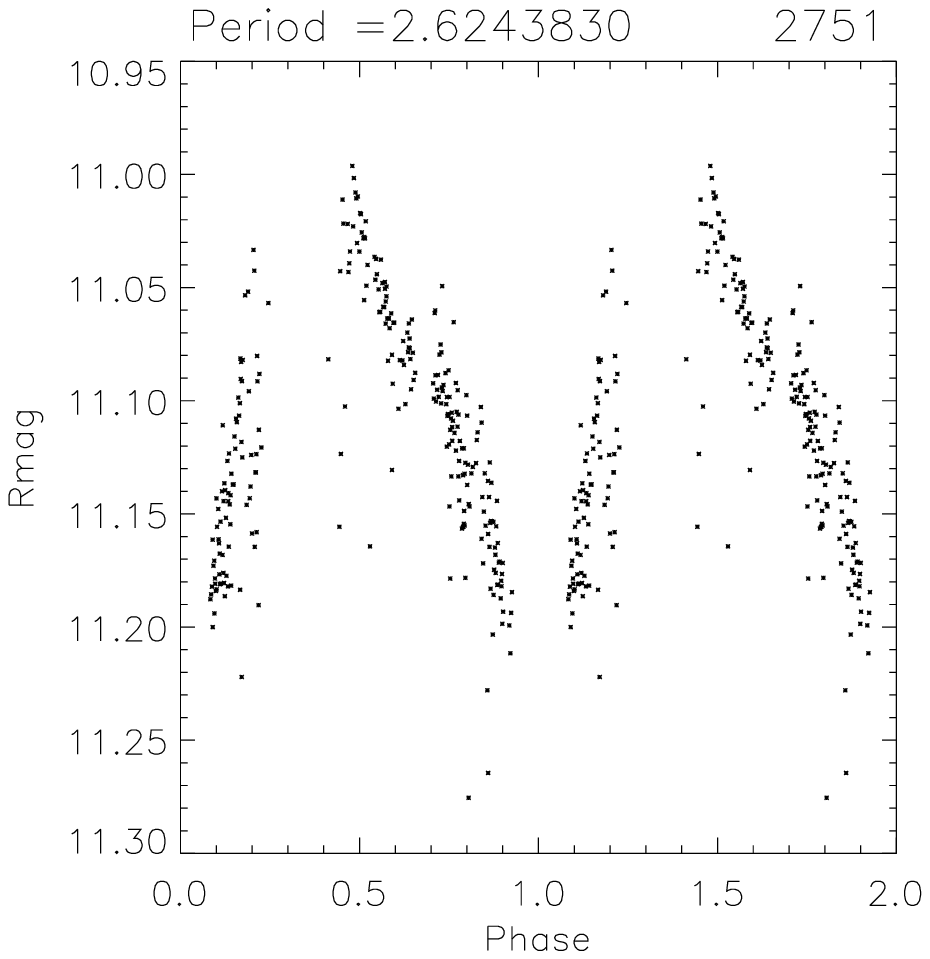}
\includegraphics[scale=.45]{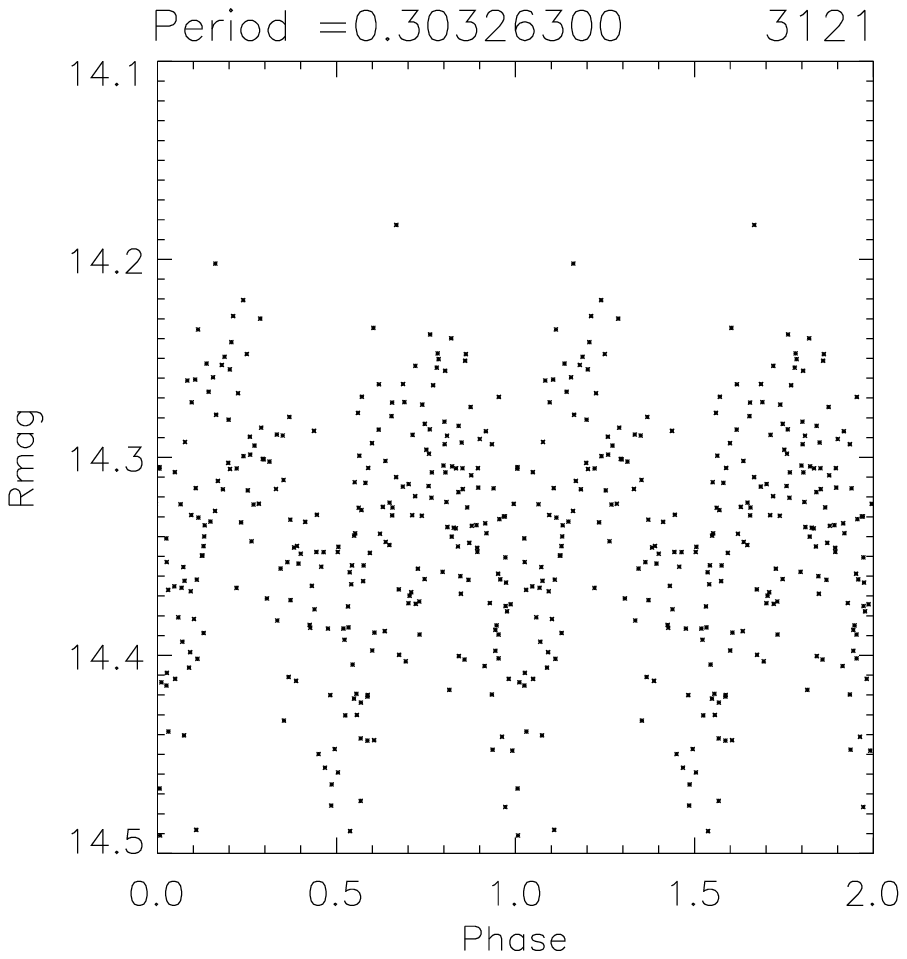}
\includegraphics[scale=.45]{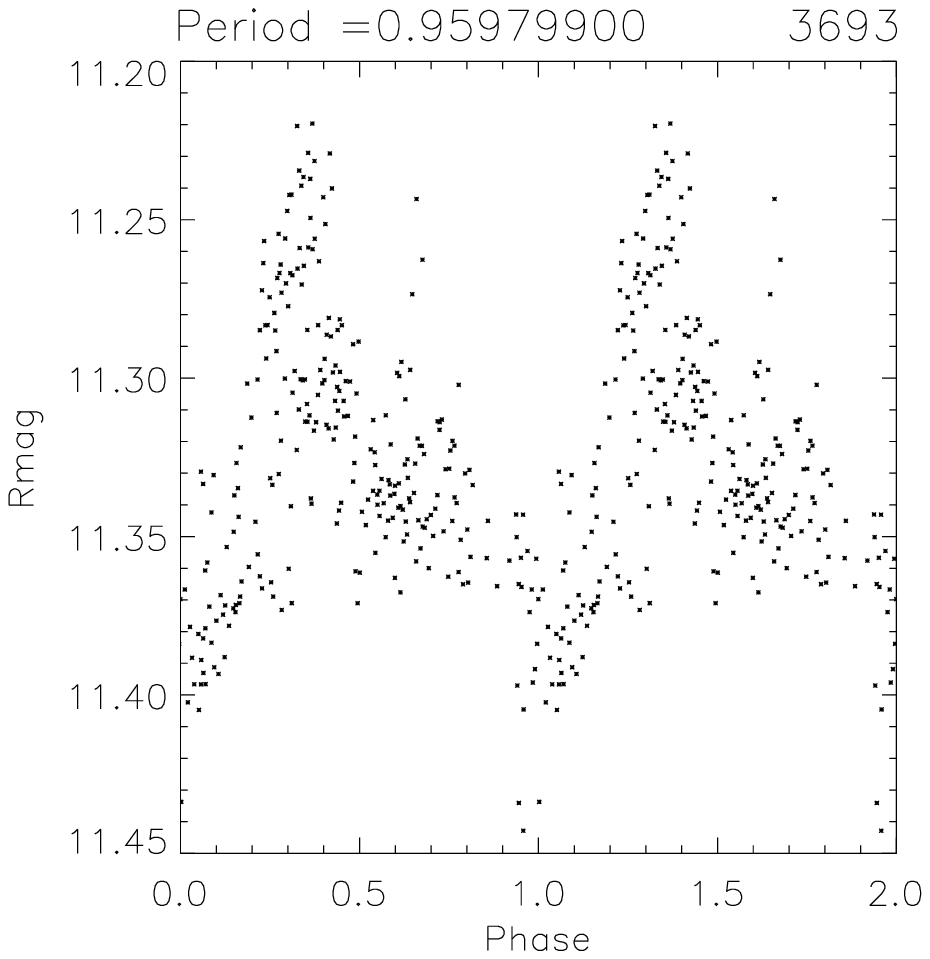}
\end{figure}
\clearpage
\begin{figure}[ht]
\includegraphics[scale=.45]{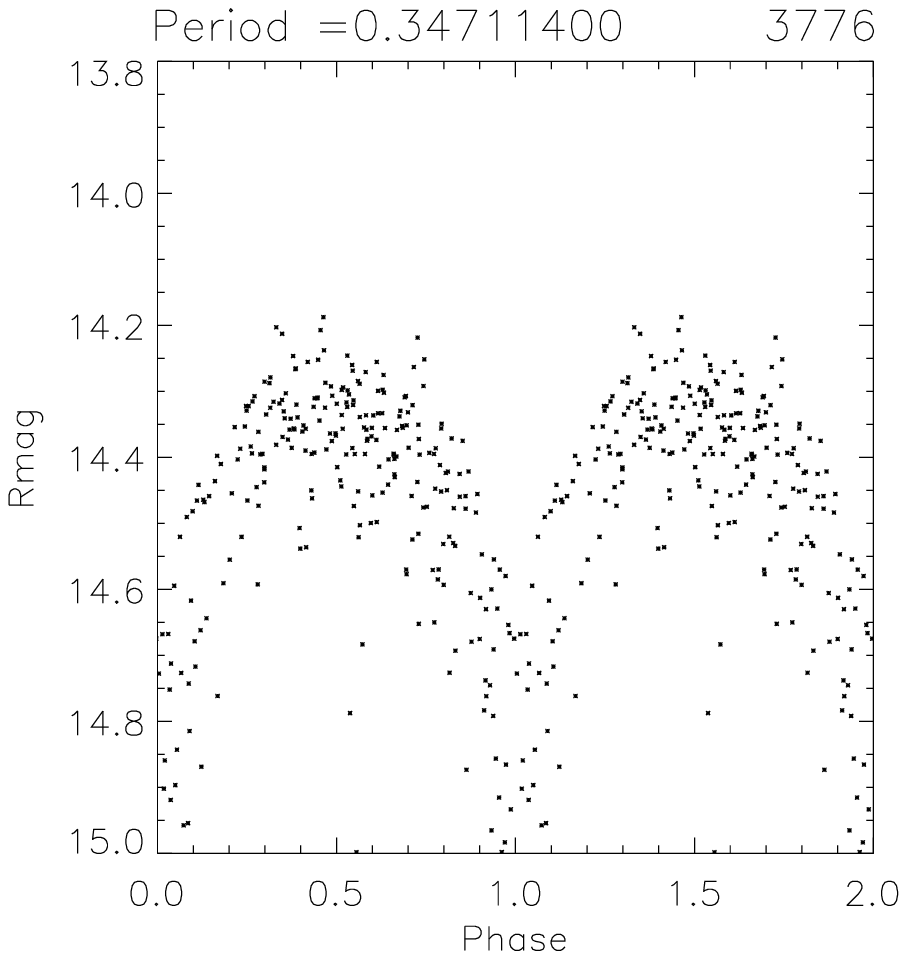}
\includegraphics[scale=.45]{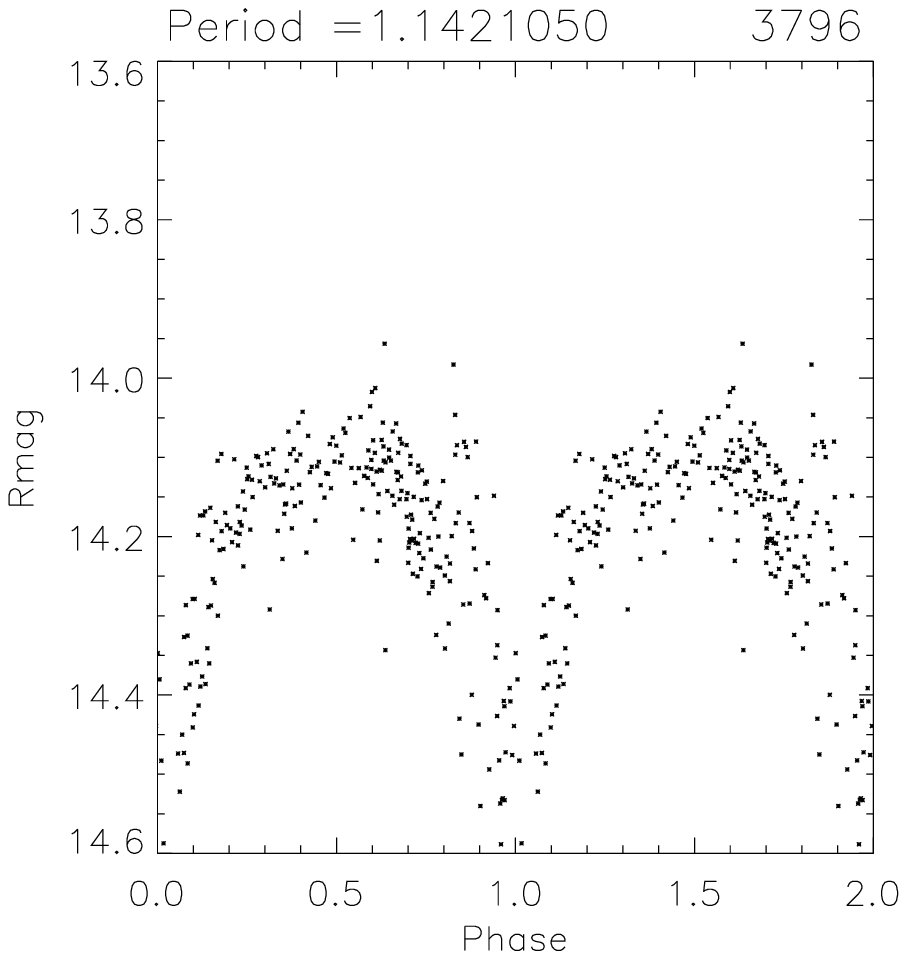}
\includegraphics[scale=.45]{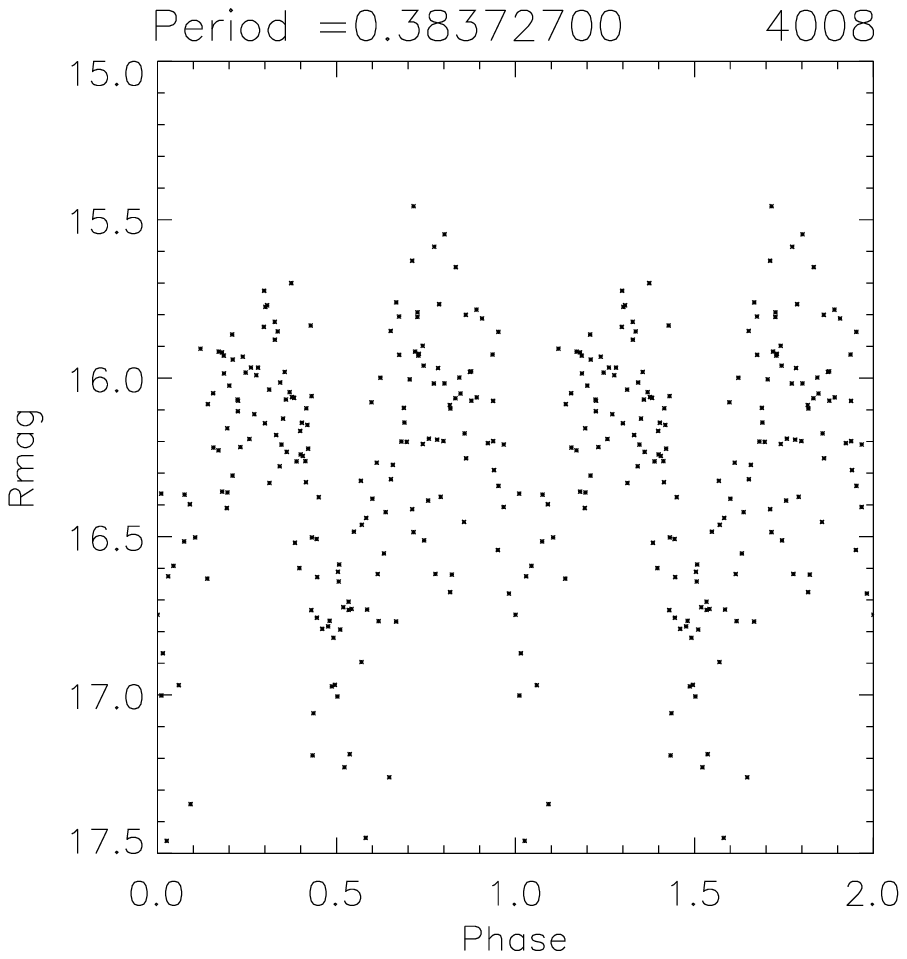}
\includegraphics[scale=.45]{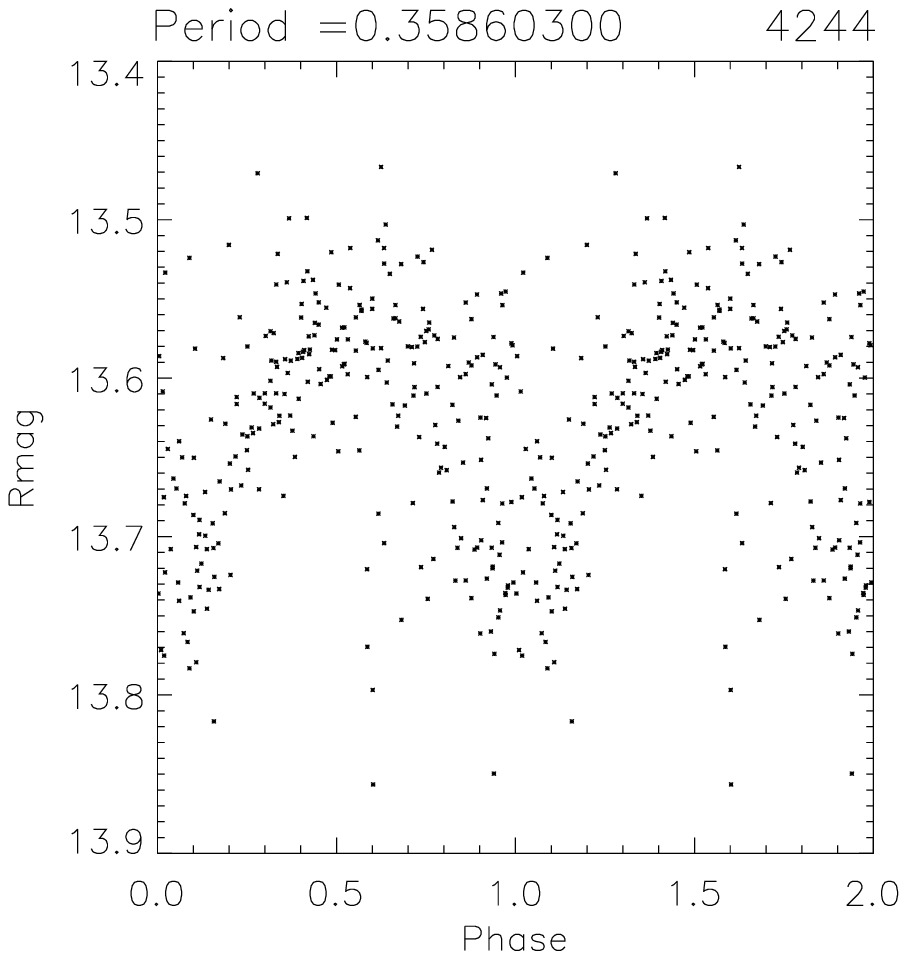}
\includegraphics[scale=.45]{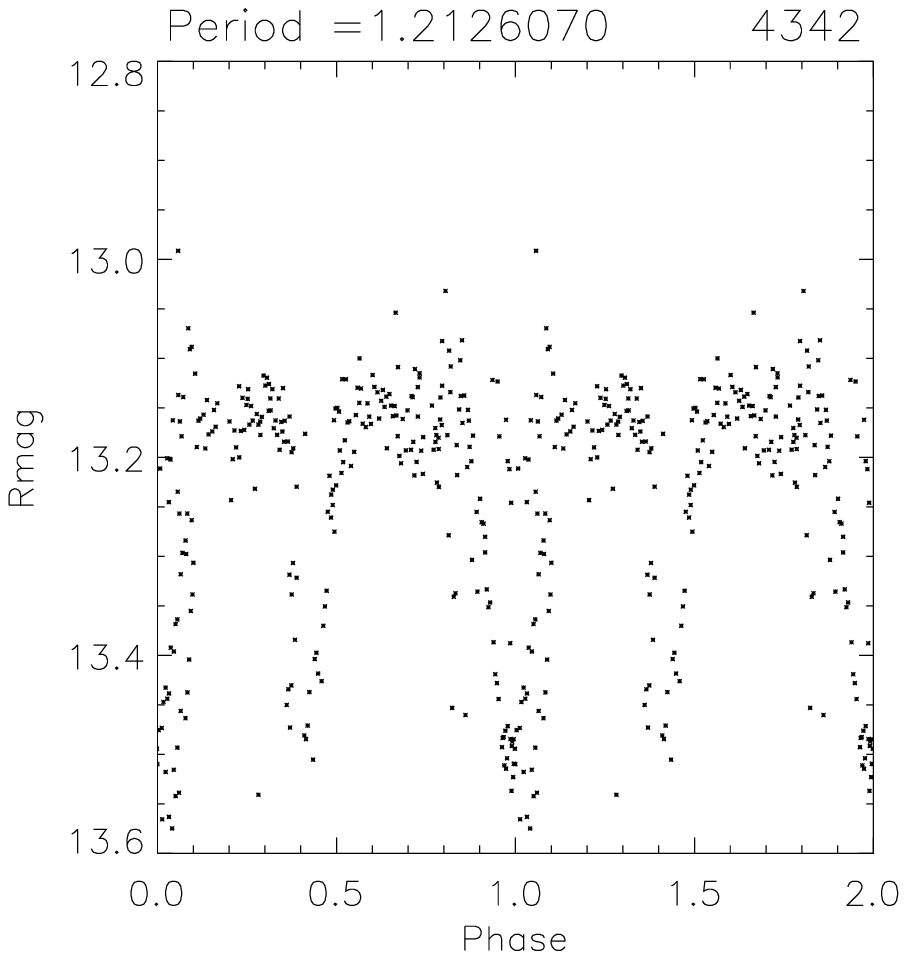}
\includegraphics[scale=.45]{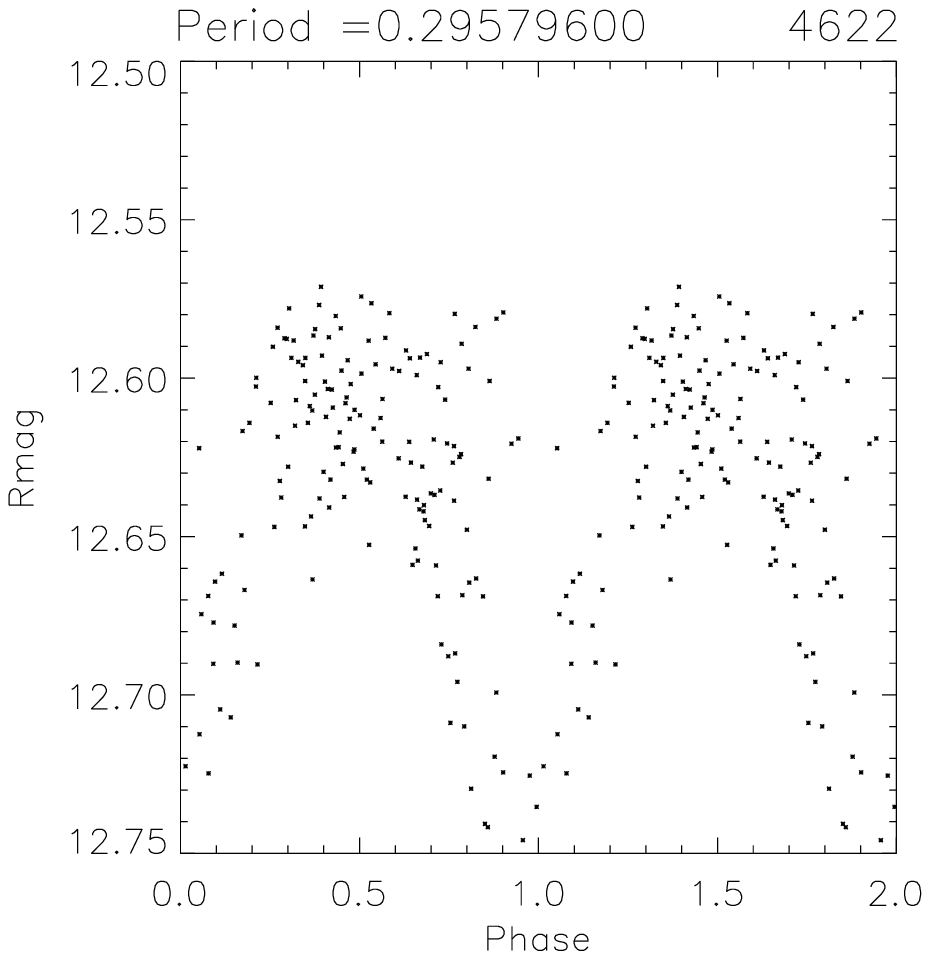}
\includegraphics[scale=.45]{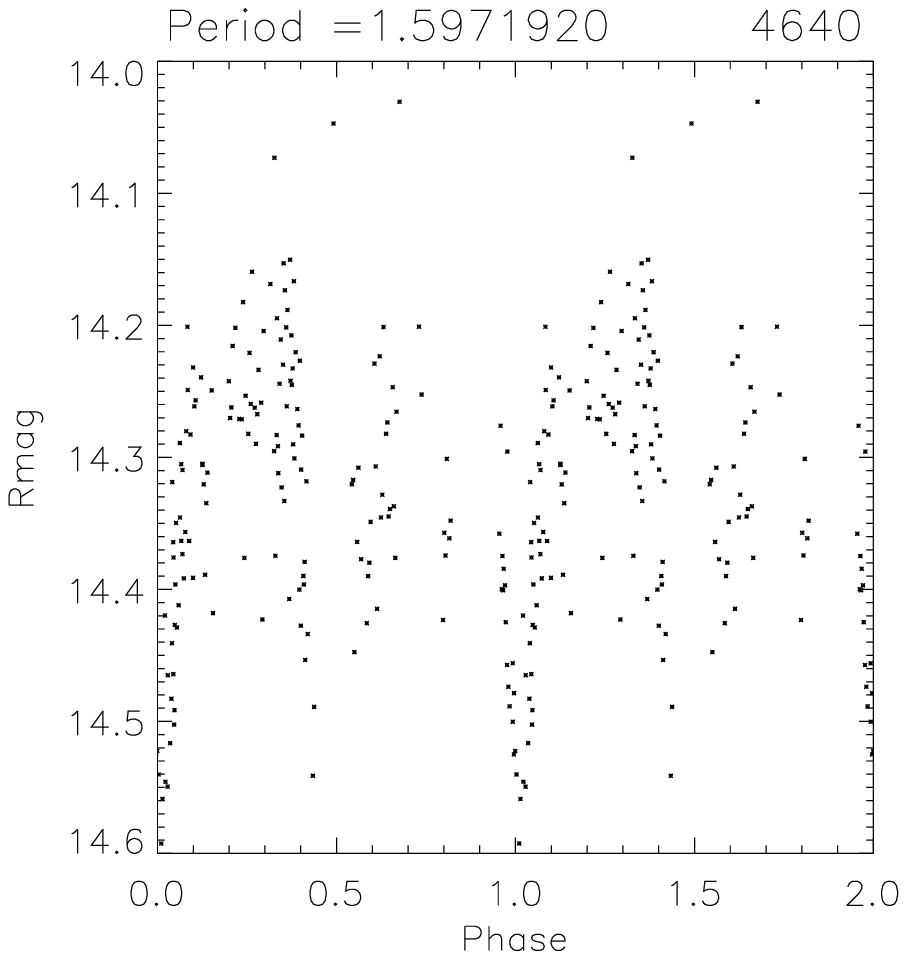}
\includegraphics[scale=.45]{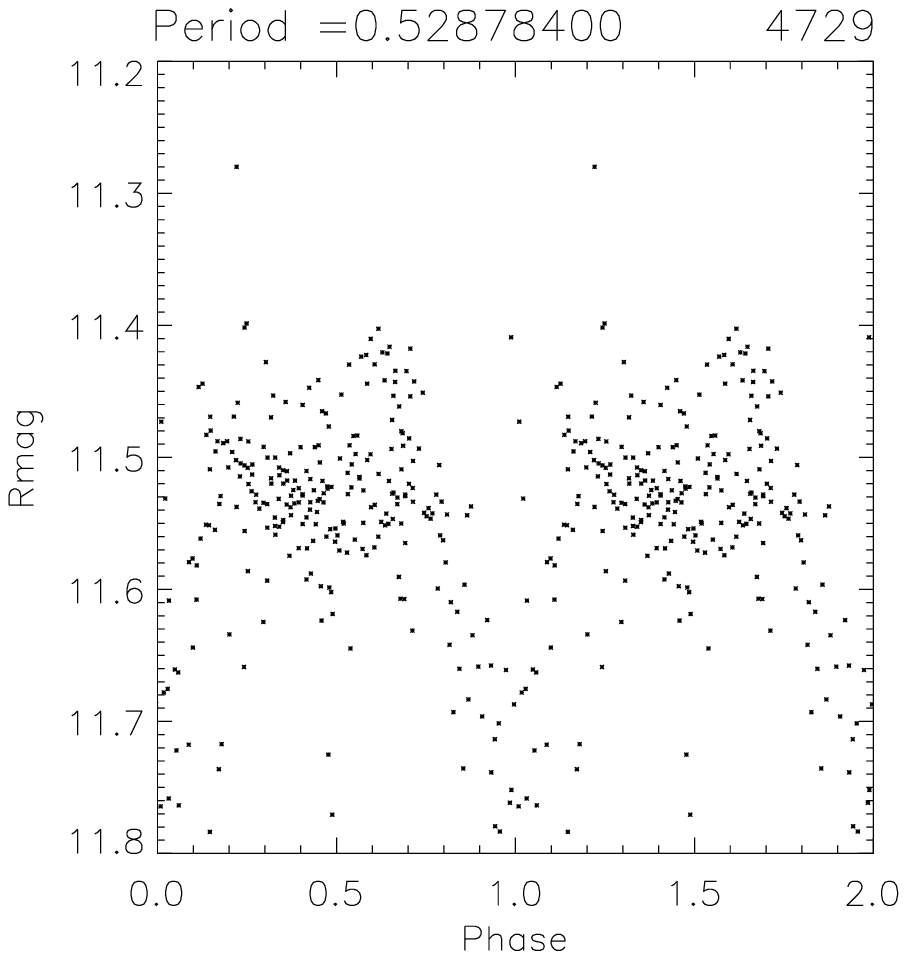}
\includegraphics[scale=.45]{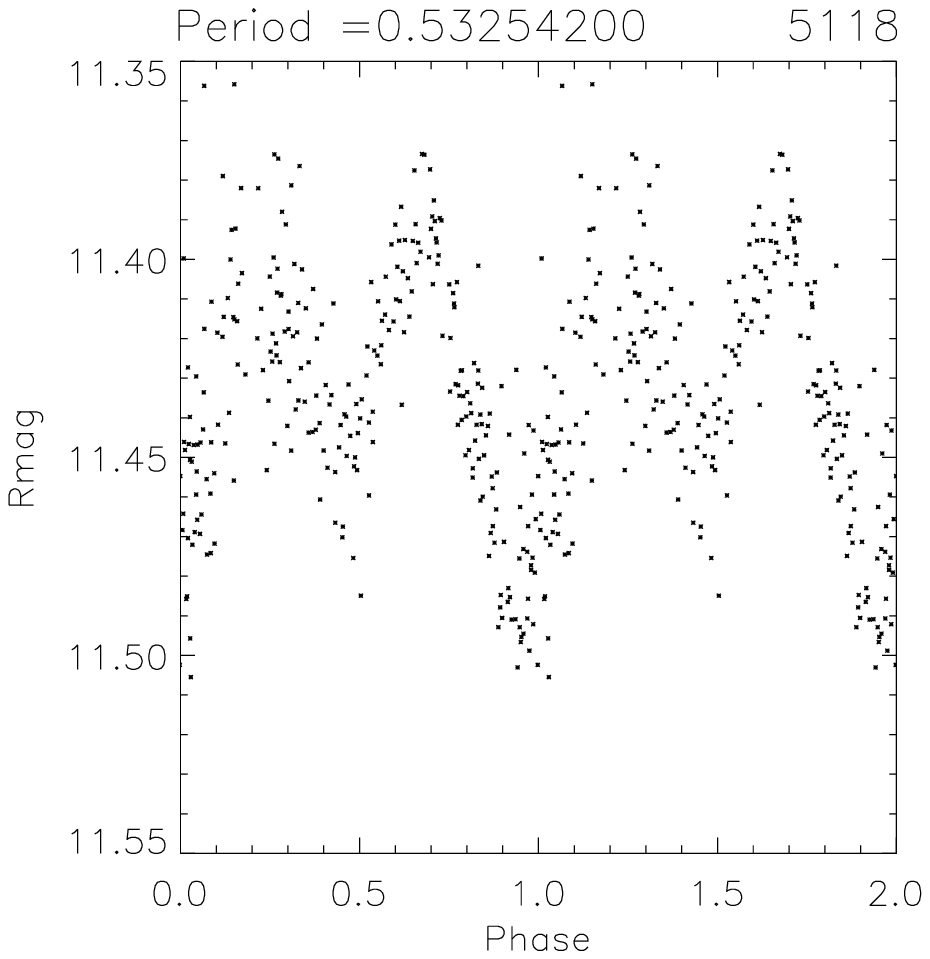}
\includegraphics[scale=.45]{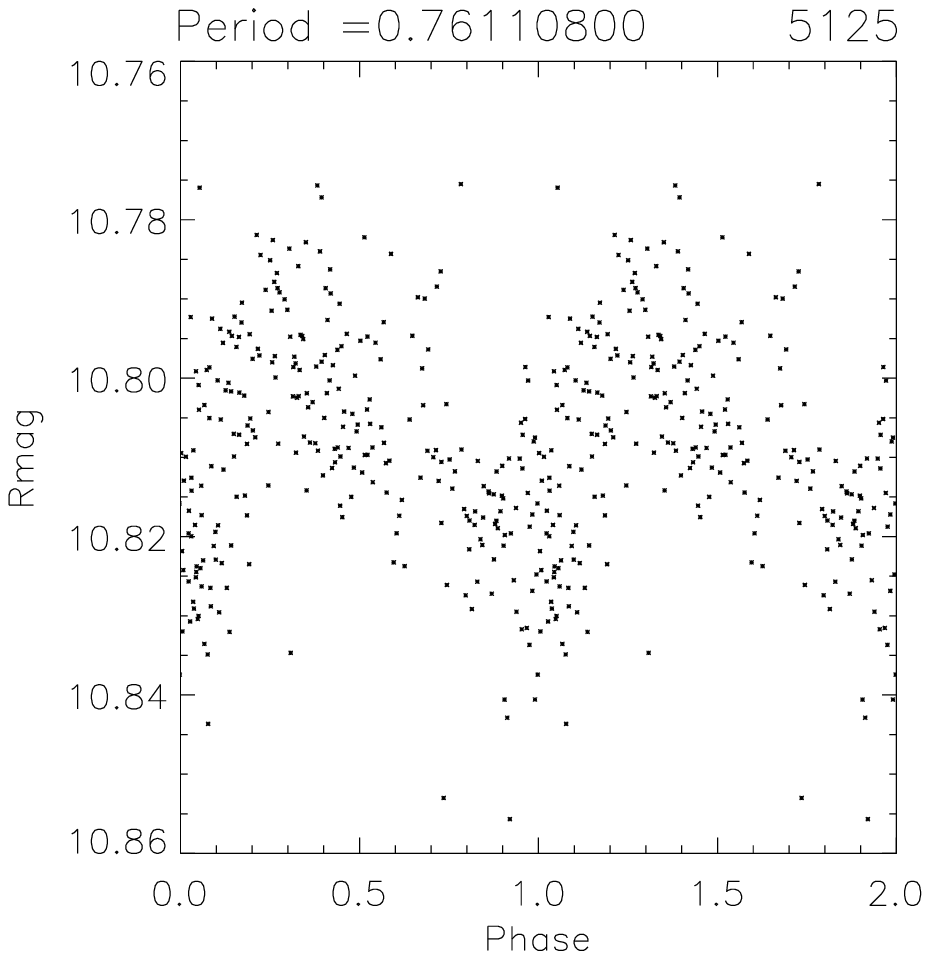}
\includegraphics[scale=.45]{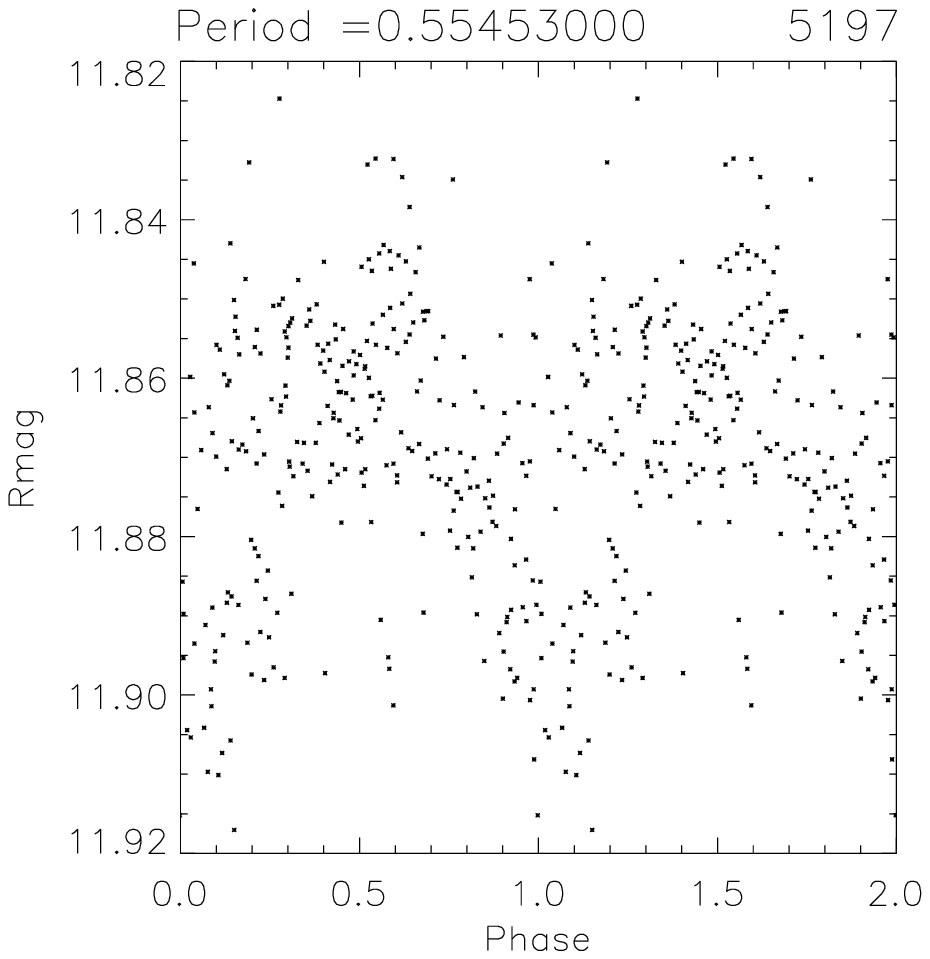}
\includegraphics[scale=.45]{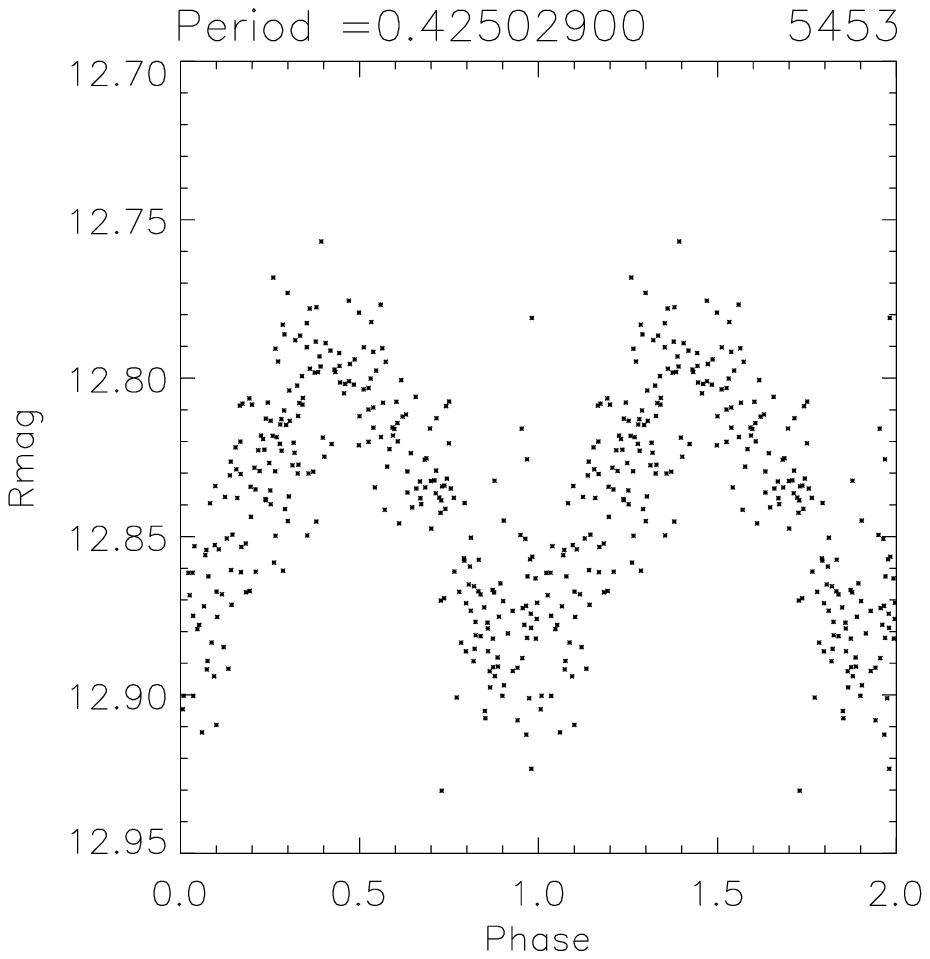}
\end{figure}
\clearpage
\begin{figure}[ht]
\includegraphics[scale=.45]{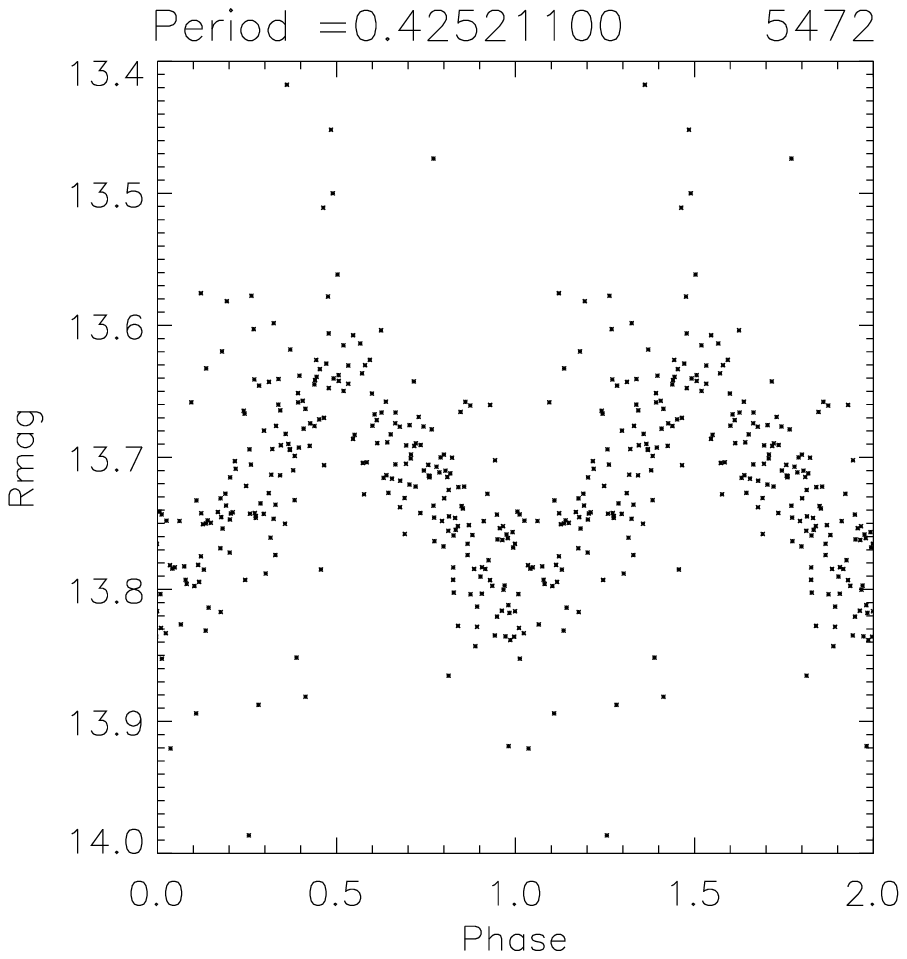}
\includegraphics[scale=.45]{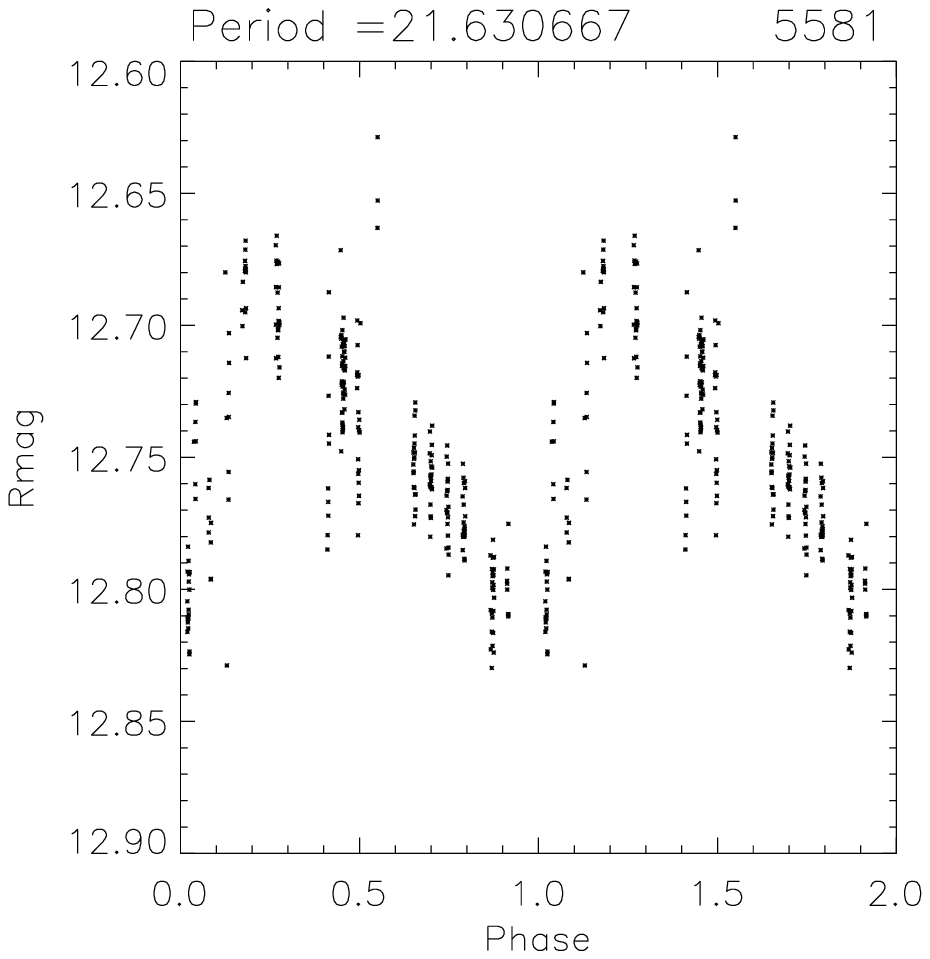}
\includegraphics[scale=.45]{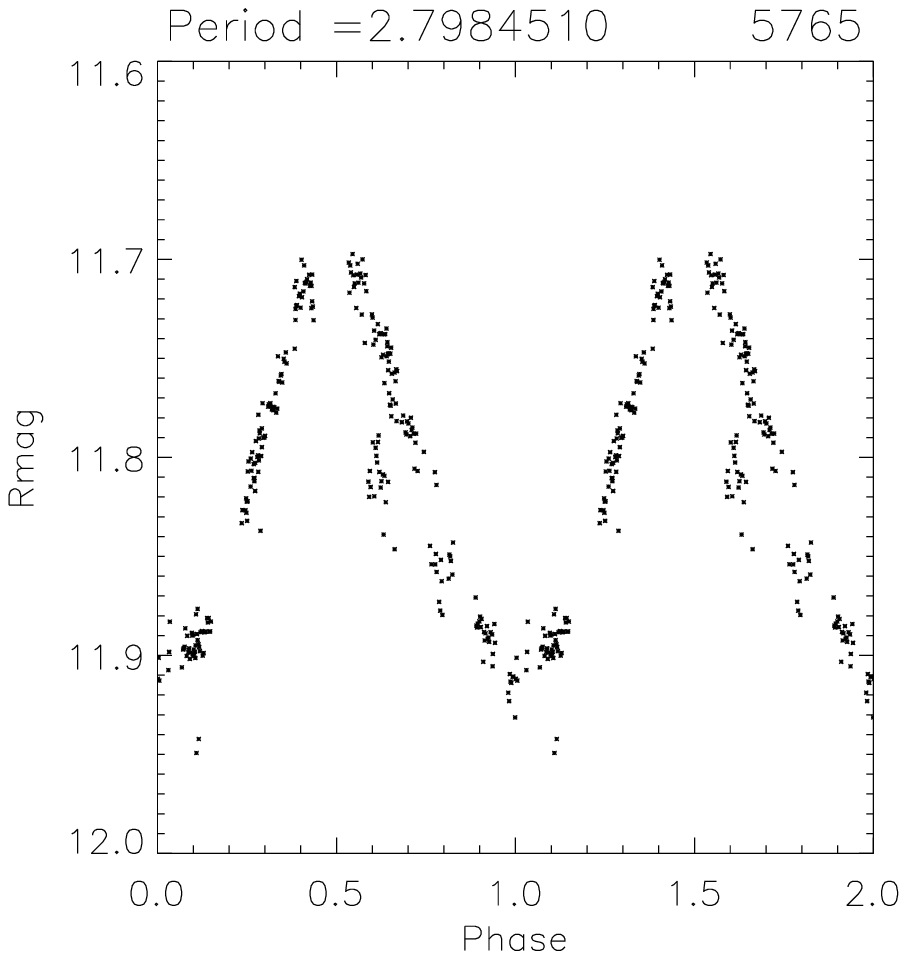}
\caption{Variable stars newly identified in the $2006$ campaign in the LRa1 field.}
\end{figure}

\clearpage

\begin{figure}[ht]
\includegraphics[scale=.45]{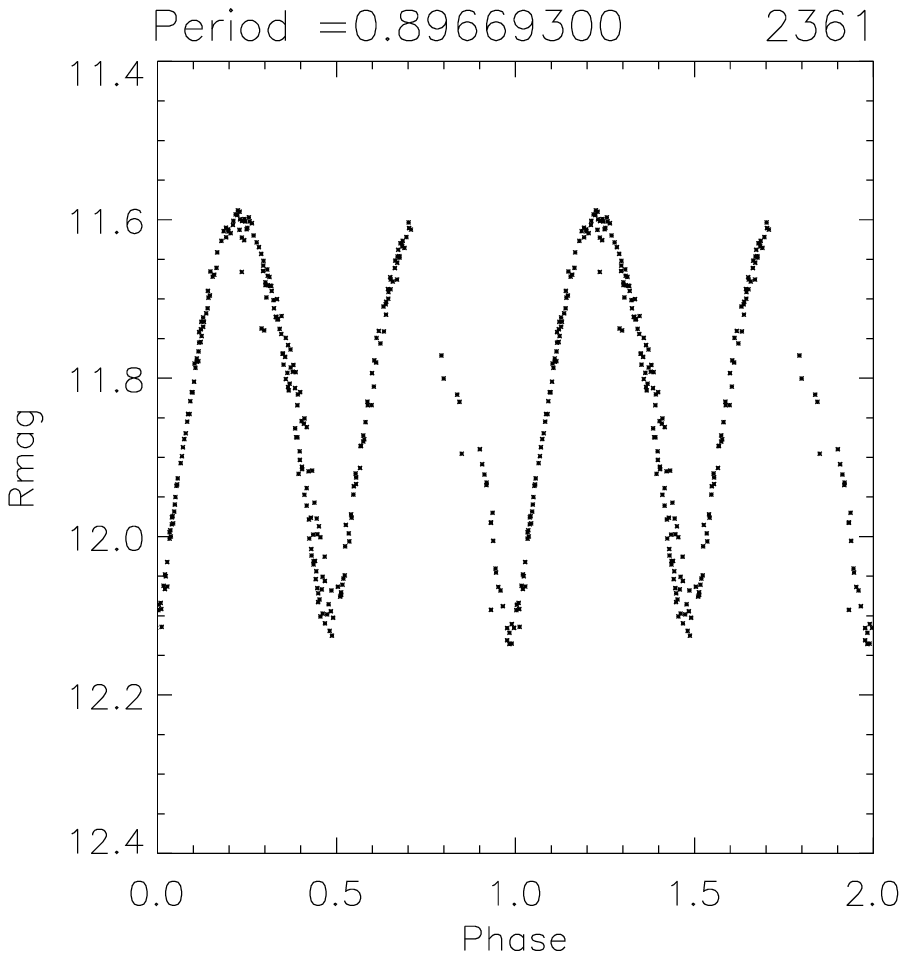}
\includegraphics[scale=.45]{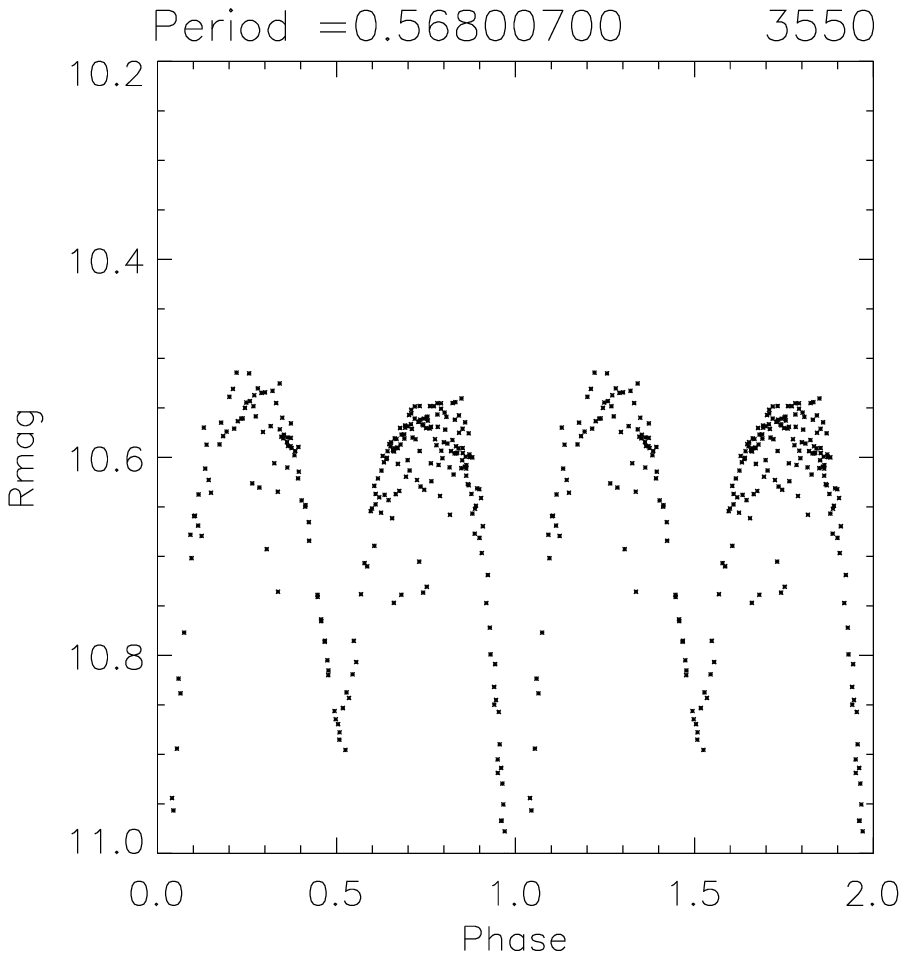}
\includegraphics[scale=.45]{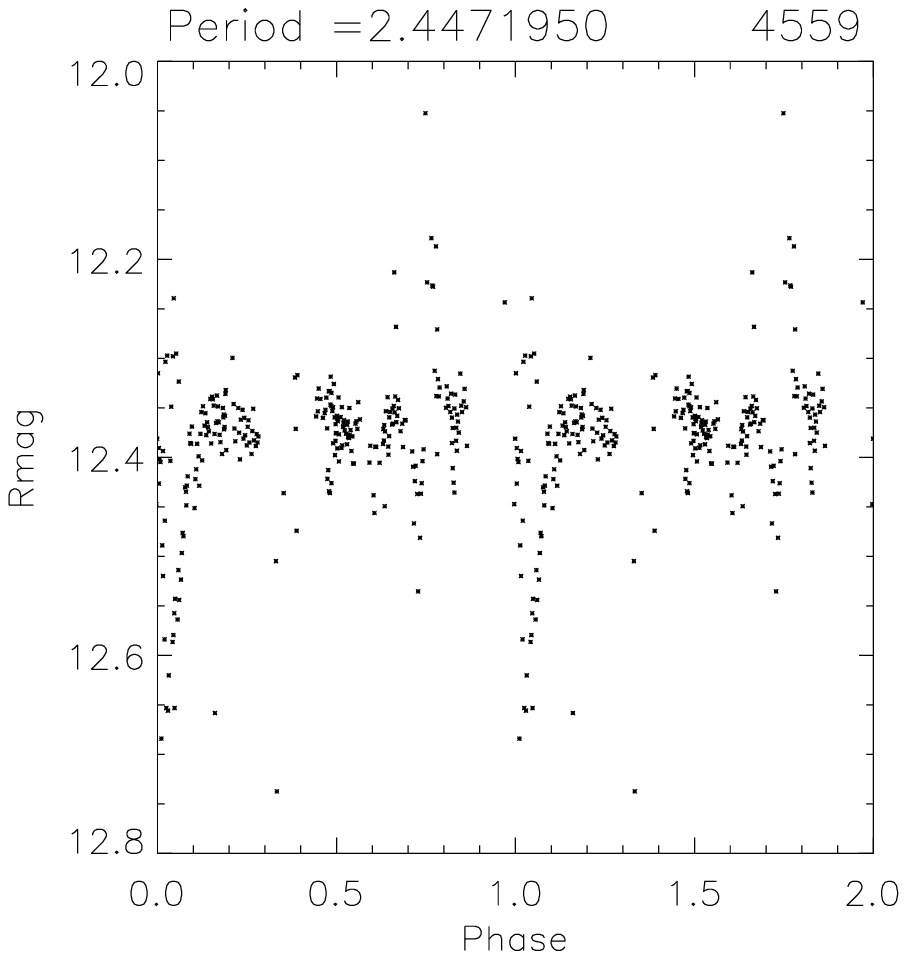}
\includegraphics[scale=.45]{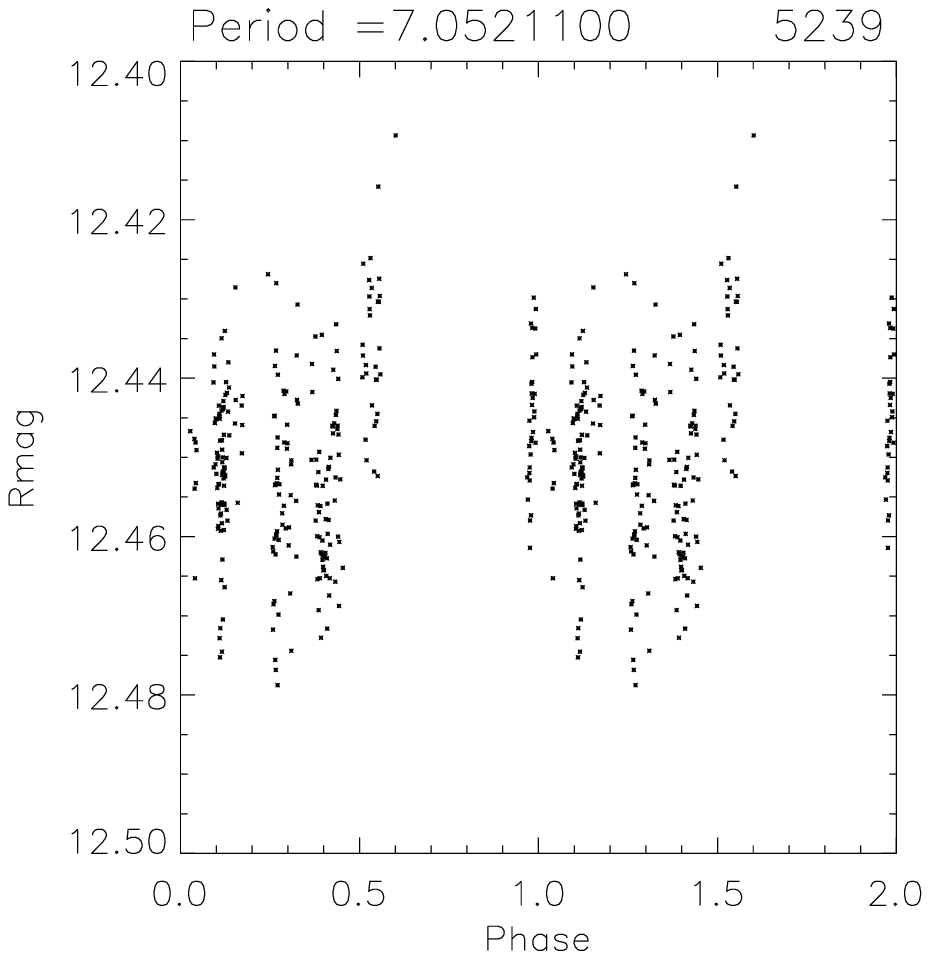}
\includegraphics[scale=.45]{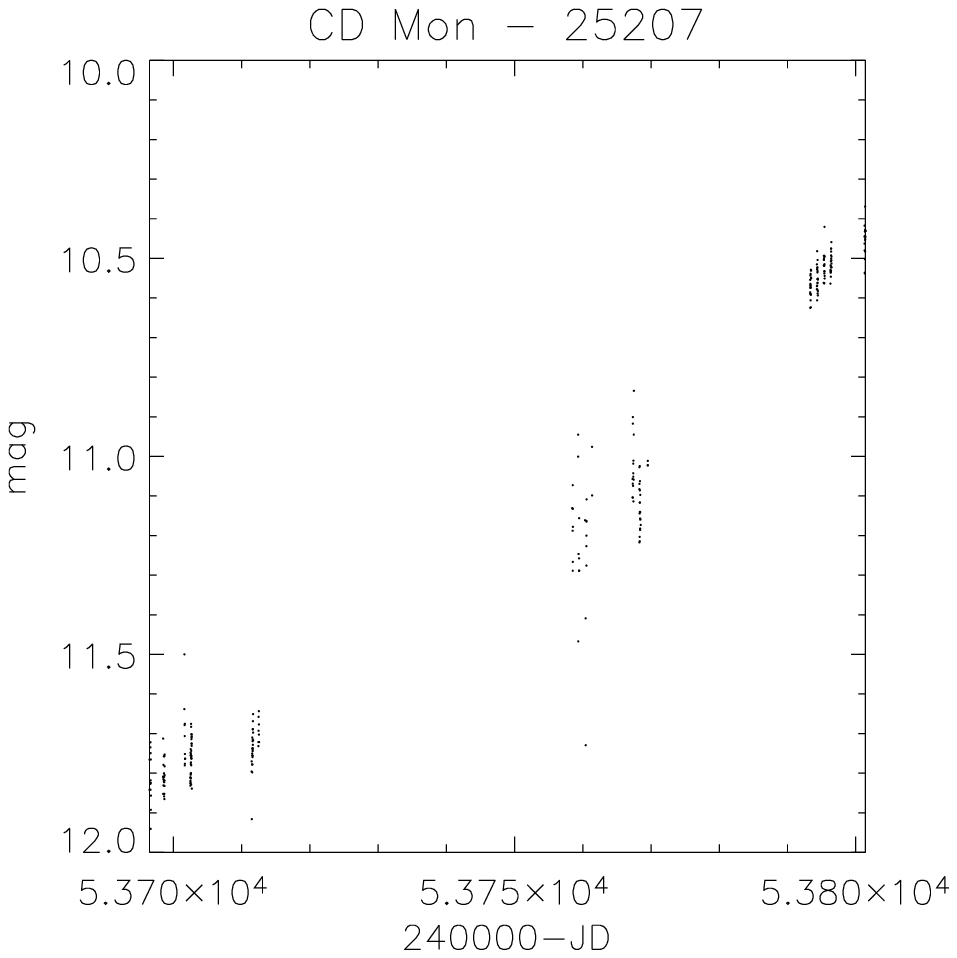}
\caption{The folded lightcurves of stars GU Mon (2361), DD Mon (3550), V 404 Mon (4559), V 501 Mon (5239) and a lightcurve of Mira-type star CD Mon (25207) are shown.}
\end{figure}

\clearpage

\begin{deluxetable}{ccccccc}

\tablecaption{Equivalent Widths}

\tablewidth{0pt}

\tablecaption{Periodic variable stars detected. Magnitudes are based on calibration against USNO catalogue only. IDs marked with asterisk are within CoRoT field of view.}

\tablehead{

\colhead{BEST ID} & 

\colhead{$\alpha$(J2000)} &

\colhead{$\delta$(J2000)} & 
\colhead{Period(days)} &

\colhead{Amplitude(mag)} &
\colhead{Mean mag} &
\colhead{Type}

}
\startdata 

      34$^{*}$ & 6 40 12 &1 9 43& 0.865& 12.815&0.170&CEP\\
      47$^{*}$ & 6 43 17 &1 9 39& 1.921& 13.035&0.420&CEP\\
      70$^{*}$ & 6 41 9 &1 9 5& 0.837& 13.250&0.200&CEP\\
     211 & 6 39 22 &1 5 28& 0.429& 10.434&0.080&CEP\\
     326$^{*}$ & 6 43 36 &1 3 1& 0.497& 14.004&0.400&CEP\\
     489$^{*}$ & 6 44 15 &0 59 58& 0.543& 13.090&0.090&EB\\
     515$^{*}$ & 6 44 15 &0 59 35& 0.213& 13.053&0.160&DSCT\\
     856$^{*}$ & 6 43 24 &0 52 48& 0.183& 14.151&0.240&DSCT\\
     935$^{*}$ & 6 44 52 &0 50 4& 1.923& 11.543&0.080&ELL\\
     995$^{*}$ & 6 43 49 &0 48 10& 0.367& 12.655&0.170&ELL\\
    1358 & 6 47 0 &0 39 14& 1.095& 12.413&0.100&ELL\\
    1835 & 6 38 14 &0 26 25& 0.550& 12.467&0.250&EW\\
    1873$^{*}$ & 6 42 26 &0 25 21& 0.275& 11.491&0.200&EW\\
    2044$^{*}$ & 6 40 5 &0 20 25& 7.976& 12.363&0.200&CEP\\
    2057 & 6 37 23 &0 20 0& 0.485& 14.418&0.500&CEP\\
    2119$^{*}$ & 6 44 41 &0 19 2& 0.289& 13.065&0.700&CEP\\
    2300$^{*}$ & 6 40 41 &0 14 26& 0.423& 11.290&0.120&CEP\\
    2311 & 6 37 13 &0 14 8& 1.295& 13.982&0.950&EB\\
    2361$^{*}$ & 6 44 47 &0 13 18& 0.897& 11.821 &0.550&GU Mon\\
    2453 & 6 35 51 &0 10 4& 0.469& 14.268&0.500&CEP\\
    2462 & 6 37 2 &0 9 48& 0.193& 12.272&0.170&DSCT\\
    2570$^{*}$ & 6 43 21 &0 7 23& 0.404& 12.701&0.080&CEP\\
    2751 & 6 39 50 &0 2 30& 2.624& 11.118&0.200&CEP\\
    3121$^{*}$ & 6 45 30 &-0 7 11& 0.303& 14.338&0.250&EW\\
    3550$^{*}$ & 6 45 58 &-0 17 31& 0.568& 10.610&0.480&DD Mon\\
    3693$^{*}$ & 6 41 50 &-0 20 27& 0.960& 11.331&0.180&CEP\\
    3776$^{*}$ & 6 45 36 &-0 22 5& 0.347& 14.414&0.800&CEP\\
    3796$^{*}$ & 6 45 36 &-0 22 33& 1.142& 14.173&0.550&EB\\
    4008 & 6 47 1 &-0 28 38& 0.384& 16.208&0.250&EW\\
    4244 & 6 39 0 &-0 34 53& 0.359& 13.612&0.270&CEP\\
    4342$^{*}$ & 6 41 52 &-0 37 40& 1.213& 13.191&0.500&EA\\
    4559 & 6 39 29 &-0 44 31&2.447& 12.378 &0.400&V404 Mon\\    
    4622$^{*}$ & 6 47 18 &-0 45 48& 0.296& 12.622&0.170&EA\\
    4640 & 6 47 21 &-0 46 19& 1.597& 14.320&0.450&EB\\
    4729 & 6 38 35 &-0 49 14& 0.529& 11.534&0.400&EA\\
    5118$^{*}$ & 6 42 46 &-1 1 8& 0.533& 11.436&0.150&ELL\\
    5125$^{*}$ & 6 40 59 &-1 1 23& 0.761& 10.809&0.060&CEP\\
    5197$^{*}$ & 6 43 36 &-1 3 37& 0.555& 11.868&0.080&CEP\\
    5239$^{*}$ & 6 45 28 &-1 4 46&7.052& 12.450 &0.070&V501 Mon\\    
    5453$^{*}$ & 6 42 16 &-1 10 52& 0.425& 12.835&0.140&CEP\\
    5472$^{*}$ & 6 42 18 &-1 11 24& 0.425& 13.722&0.250&CEP\\
    5581$^{*}$ & 6 45 36 &-1 14 19&21.631& 12.752&0.140&CEP\\
    5765$^{*}$ & 6 41 52 &-1 20 05& 2.798& 11.800&0.200&CEP\\
    25207$^{*}$ & 6 40 30 &-1 4 45&---& 14.6369 &---&CD Mon\\             
\nl
    
\enddata

\end{deluxetable} 

\end{document}